\documentclass{aastex61}

\newcommand\aastex{AAS\TeX}

\submitjournal{ApJS}

\shorttitle{\aastex\ Broadband Spectral Investigations of Magnetar Bursts}
\shortauthors{Kirmizibayrak et al.}
\begin{document}

\title{Broadband Spectral Investigations of Magnetar Bursts}
\correspondingauthor{Demet Kirmizibayrak}
\email{demetk@sabanciuniv.edu}

\author{Demet K{\i}rm{\i}z{\i}bayrak}
\affiliation{Faculty of Engineering and Natural Sciences, Sabanc{\i} University, Orhanl{\i} Tuzla, Istanbul 34956, Turkey}

\author{Sinem \c{s}a\c{s}maz Mu\c{s}}
\affiliation{Faculty of Science and Letters, Istanbul Technical University, 34469, Maslak, Istanbul, Turkey}
\affiliation{Faculty of Engineering and Natural Sciences, Sabanc{\i} University, Orhanl{\i} Tuzla, Istanbul 34956, Turkey}

\author{Yuki Kaneko}
\affiliation{Faculty of Engineering and Natural Sciences, Sabanc{\i} University, Orhanl{\i} Tuzla, Istanbul 34956, Turkey}

\author{Ersin G\"{o}\u{g}\"{u}\c{s}}
\affiliation{Faculty of Engineering and Natural Sciences, Sabanc{\i} University, Orhanl{\i} Tuzla, Istanbul 34956, Turkey}

\begin{abstract}

We present our broadband (2 - 250 keV) time-averaged spectral analysis of 388 bursts from SGR J1550$-$5418, SGR 1900+14 and SGR 1806$-$20 detected with the Rossi X-ray Timing Explorer (RXTE) here and as a database in a companion web-catalog. We find that two blackbody functions (BB+BB), sum of two modified blackbody functions (LB+LB), sum of blackbody and powerlaw functions (BB+PO) and a power law with a high energy exponential cut-off (COMPT) all provide acceptable fits at similar levels. We performed numerical simulations to constrain the best fitting model for each burst spectrum and found that $67.6 \%$ burst spectra with well-constrained parameters are better described by the Comptonized model. We also found that $64.7 \%$ of these burst spectra are better described with LB+LB model, which is employed in SGR spectral analysis for the first time here, than BB+BB and BB+PO. We found a significant positive lower bound trend on photon index, suggesting a decreasing upper bound on hardness, with respect to total flux and fluence. We compare this result with bursts observed from SGR and AXP sources and suggest the relationship is a distinctive characteristic between the two. We confirm a significant anti-correlation between burst emission area and blackbody temperature and find that it varies between the hot and cool blackbody temperatures differently than previously discussed. We expand on the interpretation of our results in the framework of strongly magnetized neutron star case. 

\end{abstract}

\keywords{magnetars: individual --- (SGR J1550$-$5418, SGR 1900+14. SGR 1806$-$20), stars: neutron, X-rays: bursts}

\section{Introduction} \label{sec:intro}

Magnetars are neutron stars whose variety of energetic radiation mechanisms are thought to be governed by the decay of their extremely strong magnetic fields (B \({\sim} 10^{14} - 10^{15}\) G; \citealt{mereghetti2015}). Emission of energetic hard X-ray bursts is the most characteristic signature of magnetar-like behavior. There are currently 29 known magnetars (23 confirmed + 6 magnetar candidates), 18 of which (whose spin has also been measured) emitted short duration (lasting only a fraction of a second) but very luminous bursts (\citealt{kaspi2017}; \citealt{turolla2015}; \citealt{manchestercatalogue}). Magnetar bursts occur sporadically on random occasions, and the total number of bursts varies from a few to hundreds during any given burst active episode of the underlying magnetar. Each burst has the potential of revealing new insights into the burst triggering and radiation emission mechanisms.    

The principle ingredient of magnetar bursts is some type of disturbance by the extremely strong magnetic fields. According to the magnetar model, the solid crust of a neutron star could fracture when extremely large magnetic pressure builds-up on it (\citealt{thompsonandduncan95}). In this view, the scale of burst energetics would be related to the size of fractured crustal site. \citealt{thompsonlyutikovkulkarni2002} suggested that the magnetospheres of these objects are globally twisted. As an alternative burst trigger mechanism, \citealt{lyutikov2003} proposed that magnetic reconnection might take place in the twisted magnetosphere of magnetars. This magnetic reconnection is accounted for energetic flares emitted from the Sun. It is important to note that whether the trigger for short magnetar bursts is crustal or magnetospheric is still unresolved.    

The observed bursts are the end products of their initial triggers. Therefore, photons radiated away as burst might not be the direct consequence of the ignition, but a number of processes in between are likely involved. In the magnetar view, the trigger mechanism leads to the formation of a trapped fireball in the magnetosphere, composed of e\(^\pm\)-pairs as well as photons (see e.g. \citealt{thompsonandduncan95}; \citealt{beloborodovthompson2007a}). Bursts are due to radiation from these trapped pair rich fireballs. Additionally, emerging radiation is expected to be modified as it propagates through strongly magnetized and highly twisted magnetosphere (\citealt{lyub2002}). It is therefore not straightforward to unfold the underlying mechanism from the burst data.  

Spectral and temporal studies on magnetar bursts are the most important probes to help distinguish mechanisms that could modify the emerging radiation of bursts. In previous spectral investigations, both thermal and non-thermal scenarios were invoked (e.g., \citealt{feroci2004}; \citealt{israel2008}; \citealt{lin2012}; \citealt{vanderhorst2012}). In the non-thermal viewpoint (often analytically expressed as a power law with an exponential cut-off), the photons emerging from the ignition region are repeatedly Compton up-scattered by the hot e\(^\pm\)-pairs present in the magnetosphere. The corona of hot electrons may emerge in the inner dynamic magnetosphere due to field line twisting (\citealt{thompsonlyutikovkulkarni2002}; \citealt{thompsonbeloborodov2005}; \citealt{beloborodovthompson2007a}; \citealt{beloborodovthompson2007b}). Such coronas are expected to be anisotropic due to the intense and likely multipolar magnetic field around the magnetar. The density and optical thickness of the corona, as well as the electron temperature set the characteristics of a spectral cut-off energy. Consequently, the peak energy parameter of the Comptonized (often labeled as COMPT) model is interpreted in relation to the electron temperature. Time integrated spectral analysis of nearly 300 bursts from SGR J1550$-$5418 result in an average power law photon index of $-$0.92 and peak energy (E$_{peak}$) is typically around 40 keV (\citealt{vanderhorst2012}).

An alternative approach to interpret magnetar burst spectra is the thermal emission due to a short-lived thermal equilibrium of electron-photon pairs, usually described with the sum of two blackbody functions (see e.g., \citealt{feroci2004}; \citealt{vanderhorst2012}). This dual blackbody scheme approximates a continuum temperature gradient due to the total energy dissipation of photons throughout the magnetosphere. The corona is expected to be hotter at low altitudes than the outer layers. Therefore, the coronal structure suggests that the high temperature blackbody component be associated with a smaller volume than the cold component. Previous studies of magnetar burst spectra with the two blackbody model yields 3-4 keV and 10-15 keV for the temperature of cold and hot blackbodies, respectively (\citealt{olive2003}; \citealt{feroci2004}; \citealt{lin2012}; \citealt{vanderhorst2012})

In this paper, we present the results of our systematic time-averaged spectral analysis of a total of 388 single-peak bursts observed with the Rossi X-ray Timing Explorer (RXTE) between 1996$-$2009 from three magnetars; SGR J1550$-$5418, SGR 1900+14 and SGR 1806$-$20. We utilized data collected with both instruments on board RXTE. Therefore, we performed our investigations in a broad energy range of 2$-$250 keV, which is the widest energy coverage used for the analysis of SGR 1806$-$20 bursts. We modeled the time-integrated burst spectra with four different photon models, including the sum of two modified blackbody models (\citealt{lyub2002}) which is employed in spectral analysis on these sources for the first time. It is important to note that this work is focused on time-averaged spectral aspects of typical short bursts, since most events analyzed are too weak and not sufficiently long to perform time-resolved spectroscopy and the signal-to-noise ratio of HEXTE data would be too low to constrain any applied models on broadband time-resolved spectroscopy. For longer events included in the analysis, the results may not completely represent the instantaneous burst properties since SGR burst spectra are known to evolve (see e.g. \citealt{israel2010}).  \\

\section{Data and Burst Selection} \label{sec:obs}

For our broadband spectral investigations, we used data collected with the RXTE mission, which was operational over \(\sim\)16 yrs from December 1995 till the end of 2011. Throughout its mission, magnetar bursts were observed by RXTE at many occasions, especially during burst active phases. SGR J1550$-$5418  bursts included in our study were sampled from 179 pointed RXTE observations that were performed between October 2008 and April 2010. SGR 1900+14 bursts were among 432 RXTE observations between June 1998 and December 2010. SGR 1806$-$20 bursts were observed between November 1996 - June 2011 with a total of 924 pointed RXTE observations. 

We used data collected from the PCA (Proportional Counter Array) and HEXTE (High Energy X-ray Timing Experiment) instruments carried on-board RXTE. PCA and HEXTE are co-aligned with the same view but operate in different energy ranges so that a broadband energy range analysis using data collected from the two instruments (between 2$-$250 keV) is possible with an overlap between 15-60 keV. 

The PCA instrument consisted of an array of five nearly identical proportional counters filled with xenon. Each unit had a collecting area of 1600 \(cm^2\) and was optimally sensitive in the energy range of 2 - 30 keV (\citealt{jahoda2006}). Magnetar burst data collected with PCA provides medium energy resolution (64 or 256 energy channels) and a superb time resolution of 1 \(\mu\)s. HEXTE consisted of two clusters each containing four NaI/CsI scintillation counters. It was sensitive in the energy range 15 - 250 keV. The time resolution of HEXTE was 8 \(\mu\)s and the total collective area of one cluster was 800 \(cm^2\).

For the identification of bursts within the data, we performed a two-step burst identification scheme from these three magnetars using RXTE/PCA observations. We first employed a signal-to-noise ratio analysis to roughly identify the time of events, then applied a Bayesian blocks algorithm provided in \citealt{scargle2003} and the procedure discussed in \citealt{lin2013} for final identification and morphological characterization of bursts (the details of the temporal investigation part will be presented elsewhere (Sasmaz Mus et al. in preparation)). In this manner, we identified 179, 432, and 924 bursts from SGR J1550$-$5418, SGR 1806$-$20, and SGR 1900+14, respectively. We note that some of these bursts were very weak, consisting of only $\sim$10 counts. Therefore, we first examined spectra of these bursts at varying intensities, and concluded that we would need at least 80 burst counts for the PCA instrument only (after background subtraction) in order to constrain crucial spectral parameters at a statistically acceptable level. Therefore, we only included single-peak bursts with more than 80 PCA counts in our analysis.

\subsection{Generation of Burst Spectra}\label{subsec:generation}

We determined the integration time intervals for burst and background spectra using PCA observations as follows. For each burst, we first generated a light curve in the 2-30 keV band with 0.125 s resolution spanning from 100 s before the peak time until 100 s after to extract background information. We defined two nominal background extraction intervals; from 80 to 5 s before the burst, and from 5 to 80 s in the post burst episode. We excluded the time intervals of other short bursts from the background spectral integration (see the bottom panel of Figure 1). We then generated a finer light curve (2 ms resolution) in the same energy interval and selected the burst spectrum time interval. We excluded the time interval(s) during which the count rate exceeded 18000 counts/s/PCU in order to avoid any pulse pile-up related issues (Figure 1, top panel). 

\begin{figure}[h]
  \centering
      \includegraphics[scale=0.5]{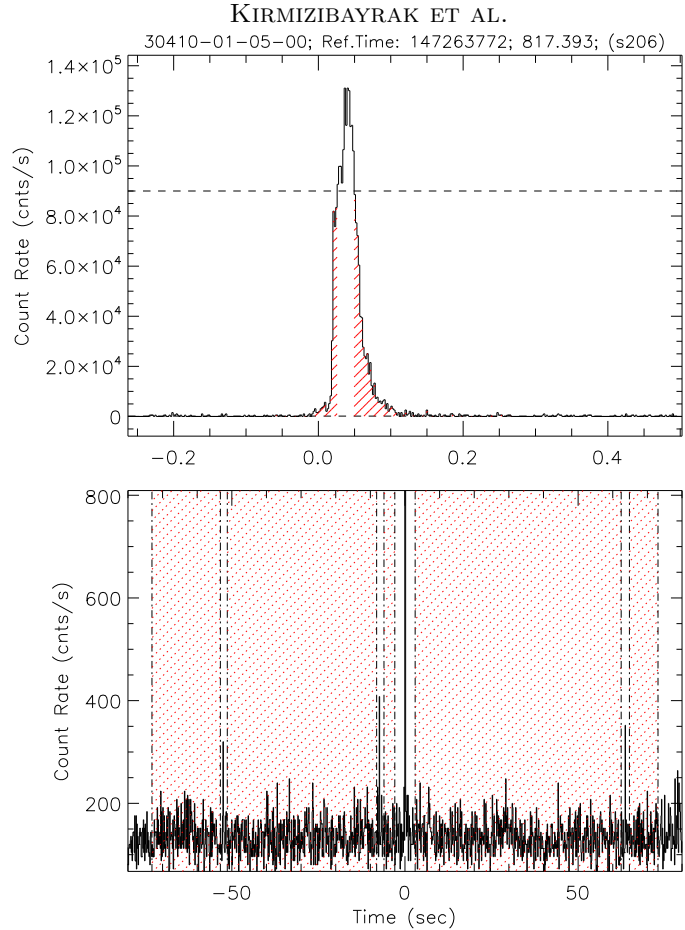}
  \caption{Burst (top figure) and background (bottom figure) selection for an SGR 1900+14 burst for spectral analysis. Red regions indicate parts used in spectral analysis. The dashed line in the top figure indicates the limit above which pulse pile-up occurs. The upper limit is 90000 counts/second since 5 PCUs were active during this observation. The time resolutions are 2 ms (top) and 0.125 s (bottom)}
\end{figure}

We used the time intervals obtained from our PCA data to generate HEXTE source and background spectra. At this point, we excluded the bursts that happened while one of the two HEXTE clusters was in "rocking mode", switching to a different direction to obtain background emission, or when no associated HEXTE data were available. For the remaining bursts, we combined the spectra obtained from both clusters. When only one of the clusters were operating during an observation, we extracted spectra using data collected with that particular cluster only. Finally, we grouped the extracted spectra of PCA and HEXTE to ensure that each spectral bin to contain a minimum of 20 burst counts.

As a result of the aforementioned eliminations, the final numbers of bursts included in our spectral analysis are: 42 for SGR J1550$-$5418, 125 for SGR 1900+14 and 221 for SGR 1806$-$20. In Tables 1 through 3, we present the list of these bursts individually from these three magnetars, together with the durations of extracted spectra with saturated parts excluded ($T_{Exp}$). Bursts included have an average duration of 0.46 s ($\sigma$ =  0.28) for SGR J1550-5418, 0.46 s ($\sigma$ =  0.3) for SGR 1900+14 and 0.72 s ($\sigma$ =  0.72) for SGR 1806-20.

\begin{deluxetable}{ccccc}\label{tab:obs1}
\tablecaption{Burst times and Durations of SGR J1550$-$5418 Bursts Included in Our Analysis \label{tab:bursts1_1550}}
\tablewidth{0pt}
\tablehead{
\colhead{Burst} & \colhead{Start time} & \colhead{Start time} & \colhead{$T_{\rm Exp}$} &
\colhead{ObsID} \\
\colhead{ID} & \colhead{in MET} &  \colhead{in UTC} &  \colhead{s} &  \\
}
\startdata
1 & 475274776.845 & 2009-01-22T20:46:14 & 0.197 & 93017-10-17-00 \\ 
2 & 475275167.175 & 2009-01-22T20:52:44 & 0.445 & 93017-10-17-00 \\ 
3 & 475276110.459 & 2009-01-22T21:08:27 & 0.189 & 93017-10-17-00 \\ 
4* & 475276179.299 & 2009-01-22T21:09:36 & 0.395 & 93017-10-17-00 \\ 
5 & 475276430.580 & 2009-01-22T21:13:47 & 0.447 & 93017-10-17-00 \\ 
\enddata
\tablecomments{Table 1 is available in the machine-readable format in full. The first 5 of 42 data lines are presented here to demonstrate its form.}
\tablenotetext{1}{Burst IDs of saturated bursts are marked with asteriks.}
\tablenotetext{2}{{$T_{\rm Exp}$} refers the duration of spectral extraction interval.}
\end{deluxetable}

\begin{deluxetable}{ccccc}\label{tab:obs2}
\tablecaption{Burst times and Durations of SGR 1900+14 Bursts Included in Our Analysis \label{tab:bursts1_1900}}
\tablewidth{0pt}
\tablehead{
\colhead{Burst} & \colhead{Start time} & \colhead{Start time} & \colhead{$T_{\rm Exp}$} &
\colhead{ObsID} \\
\colhead{ID} & \colhead{in MET} &  \colhead{in UTC} & \colhead{s} &  \\
}
\startdata
1 & 139434279.341 & 1998-06-02T19:44:42 & 0.213 & 30197-02-01-00 \\ 
2 & 139439198.455 & 1998-06-02T21:06:44 & 0.291 & 30197-02-01-00 \\ 
3 & 139607213.183 & 1998-06-04T19:46:59 & 0.602 & 30197-02-01-03 \\ 
4* & 146928739.386 & 1998-08-28T13:32:25 & 0.364 & 30197-02-03-00 \\  
5 & 146929244.714 &  1998-08-28T13:40:50 & 0.471 & 30197-02-03-00 \\
\enddata
\tablecomments{Table 2 is available in the machine-readable format in full. The first 5 of 125 data lines are presented here to demonstrate its form.}
\tablenotetext{1}{Burst IDs of saturated bursts are marked with asteriks.}
\tablenotetext{2}{{$T_{\rm Exp}$} refers the duration of spectral extraction interval.}
\end{deluxetable}

\begin{deluxetable}{ccccc}\label{tab:obs3}
\tablecaption{Burst times and Durations of SGR 1806$-$20 Included in Our Analysis \label{tab:bursts1_1806}}
\tablewidth{0pt}
\tablehead{
\colhead{Burst} & \colhead{Start time} & \colhead{Start time} & \colhead{$T_{\rm Exp}$} &
\colhead{ObsID} \\
\colhead{ID} & \colhead{in MET} &  \colhead{in UTC} & \colhead{s} &  \\
}
\startdata
1* & 168976265.693 & 1999-05-10T17:51:05 & 0.395 & 40130-04-13-00 \\ 
2 & 173045819.275 & 1999-06-26T20:16:59 & 0.148 & 40130-04-20-00 \\ 
3* & 208723242.666 & 2000-08-12T18:40:42 & 0.250 & 50142-01-33-00 \\ 
4 & 212193703.626 & 2000-09-21T22:41:43 & 0.357 & 50142-01-43-00 \\ 
5* & 212194516.810 & 2000-09-21T22:55:16 & 4.154 & 50142-01-43-00 \\ 
\enddata
\tablecomments{Table 3 is available in the machine-readable format in full. The first 5 of 221 data lines are presented here to demonstrate its form.}
\tablenotetext{1}{Burst IDs of saturated bursts are marked with asteriks.}
\tablenotetext{2}{{$T_{\rm Exp}$} refers the duration of spectral extraction interval.}
\end{deluxetable}

\section{Spectral Analysis} \label{sec:spec}

In our broadband spectral analysis, we used four models, three of which have been commonly used in describing short magnetar bursts in previous studies: the sum of two blackbody functions (BB+BB), sum of blackbody and power law models (BB+PO) and Comptonized model (COMPT). Additionally, we employed the sum of two modified blackbody functions (LB + LB) as set forth by \citealt{lyub2002}. Note that the COMPT model is simply a power law with a high energy exponential cut-off expressed as:

\begin{equation}
f = AE^{-\alpha}\exp{-E/E_{cut}}
\end{equation}
where, \textit{f} is the photon flux and \textit{A} is the amplitude in photons$/$cm$^{2}/$s$/$keV at 1 keV, \(E_{cut}\) is the cut-off energy (in keV) and \(\alpha\) is the photon index. The LB function is a modified version of the blackbody function where the spectrum is flattened at low energies. In terms of the photon flux, the function is expressed as: 
\begin{equation}
f = 0.47\epsilon^2\Bigg[exp\Bigg(\frac{\epsilon^2}{T_b\sqrt{\epsilon^2+(3\pi^2/5)T_b^2}}\Bigg)-1\Bigg]^{-1}
\end{equation}
where, \(T_b\) is the bolometric blackbody temperature (temperature of a blackbody which would have its mean frequency at the peak of the observed spectrum) in keV and \(\epsilon\) is the photon energy (\citealt{lyub2002}). To display intrinsic differences of these models, we present in Figure 2 the best fit model spectra generated with the fitted parameters for the event with Burst ID 79 observed from SGR 1806$-$20. In Figure 3, we present the broadband spectrum of the same burst along with the fit residuals of all these four models.
   
  \begin{figure}[h]\label{fig:models}
\plotone{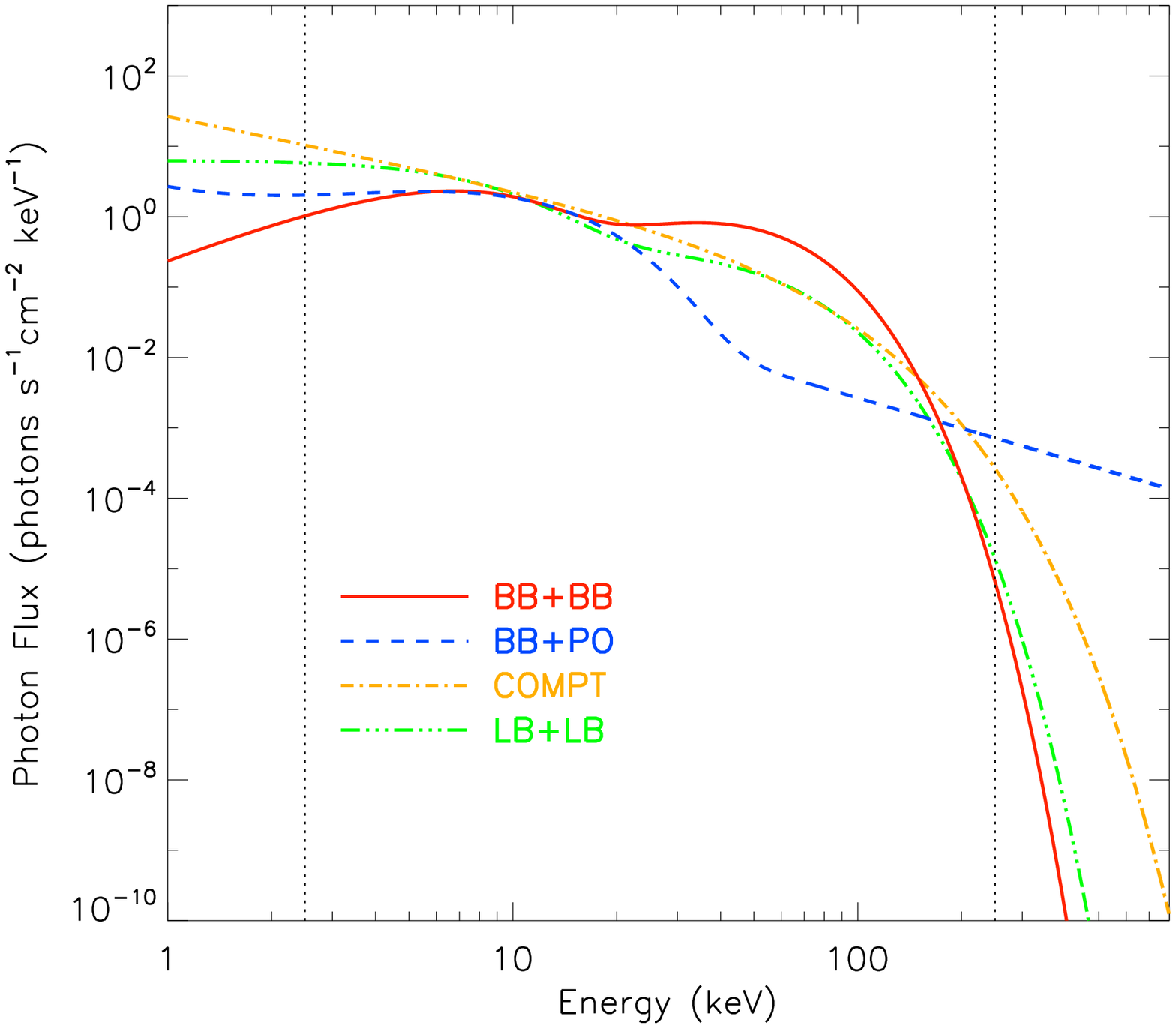}
  \caption{Comparison of spectral continuum models employed in our investigations. Model curves were obtained by fitting the spectrum of a burst from SGR 1806$-$20 (Burst ID: 79). The vertical dotted lines denote the energy interval of our spectral investigations.}
\end{figure}

 \begin{figure}[h]\label{fig:fitex}
  \centering
\includegraphics[height= 0.5\textwidth]{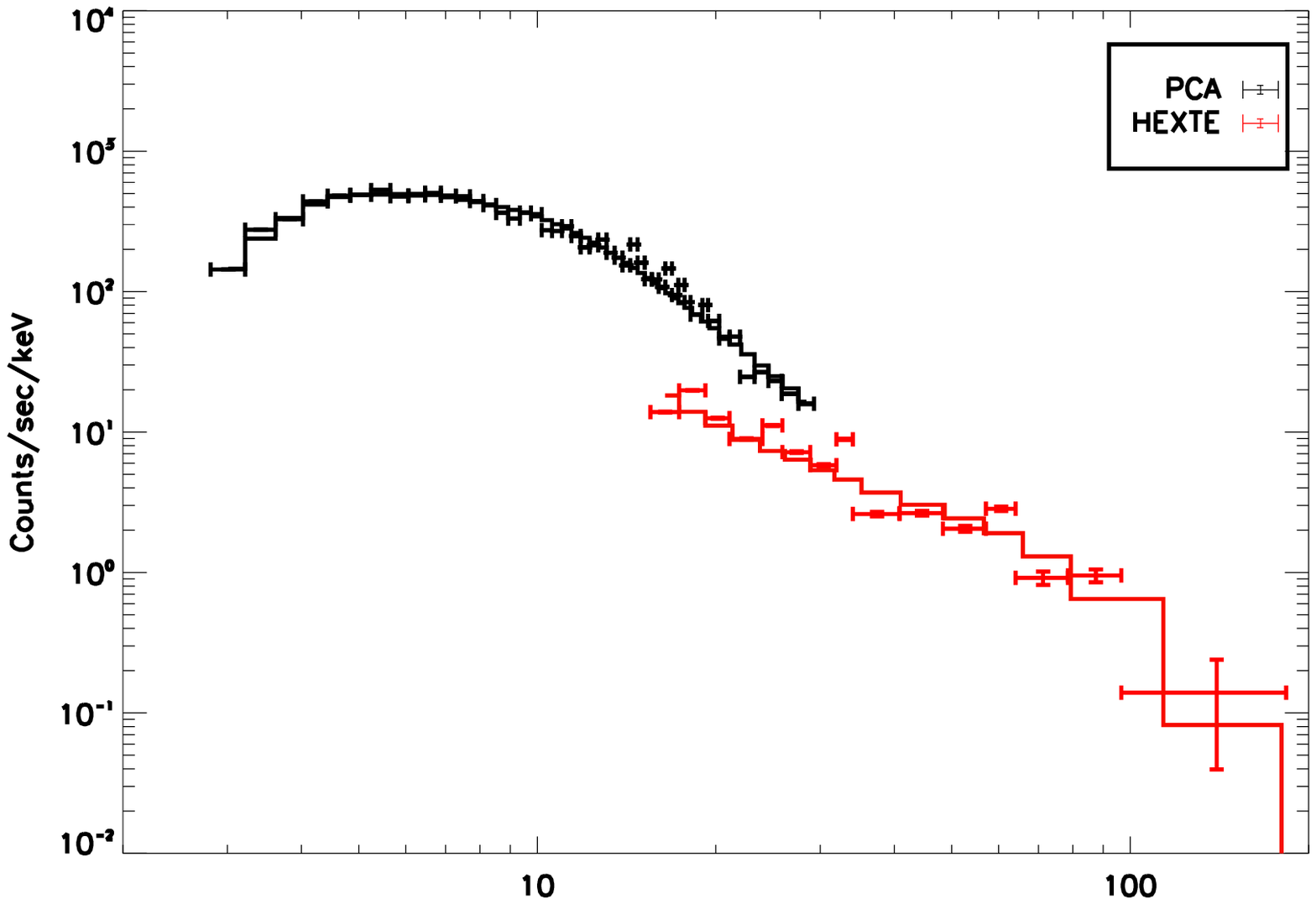}
\hspace*{-0.89cm} \includegraphics[width=0.785\textwidth, height= 0.5\textwidth]{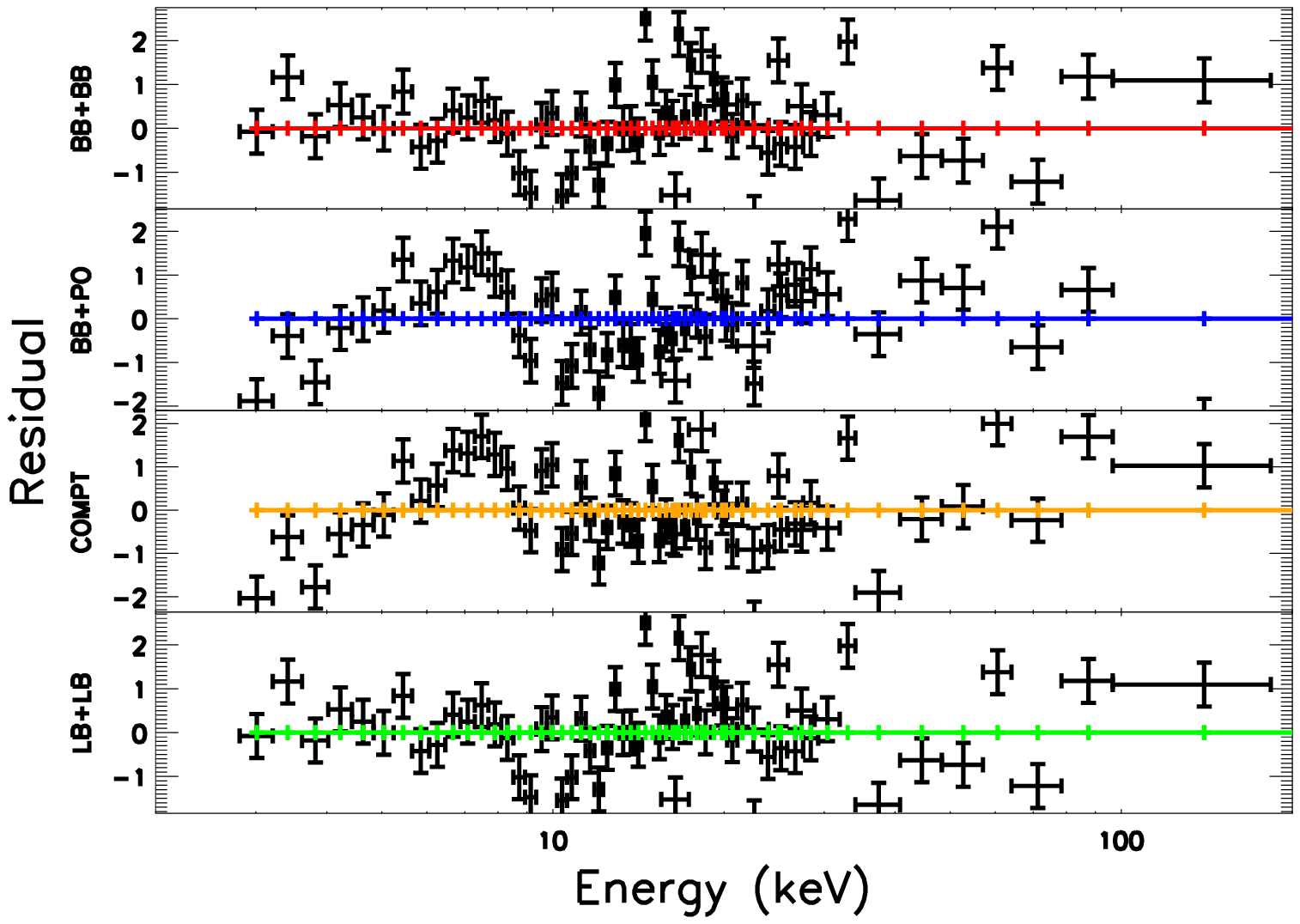}
  \caption{\textit{Top panel:} Fitted models for SGR1806$-$20 Burst ID 79.  \textit{Lower group of panels:} Fit residuals for the same event. The models for fit residuals are BB+BB, BB+PO, COMPT, and LB+LB respectively from top to bottom.}
\end{figure}
  
Before performing the joint analysis, we first investigated possible cross-calibration discrepancies between PCA and HEXTE detector responses using a small sample (11) of bursts of various flux levels. In this task, we first introduced a multiplicative constant for HEXTE parameters to account for possible discrepancies. We repeated the same analysis with the same burst sample without this scaling term. We found that the spectral analysis results with and without the constant term are in agreement with each other within 1 $\sigma$ errors for all these bursts. Therefore, we concluded that the constant term was not needed in the analysis, and proceeded our investigations without including the term to limit the number of fit parameters.

Finally, as mentioned above, we used a power law with an exponential cut-off model (COMPT) to represent the non-thermal emission spectrum. However, a different parametrization of the same model has been used in previous studies (see e.g. \citealt{lin2012}; \citealt{vanderhorst2012}; \citealt{feroci2004}) expressed as:

\begin{equation} 
f = Aexp[-E(2+\lambda)/E_{peak}](E/E_{piv})^{\lambda}
\end{equation}
where, \textit{f} is the photon flux in photons$/$cm$^{2}/$s$/$keV, \textit{A} is the amplitude with units same as f, \(E_{peak}\) is the energy (in keV) at which the spectral distribution function makes its peak in $\nu F_{\nu}$ representation, \(\lambda\) is the photon index (defined as $-\alpha$ where $\alpha$ is the photon index of the COMPT model), and \(E_{piv}\) is the pivot energy fixed at a certain value (20 keV, \citealt{lin2012}). It is then possible to convert the cut-off energy of the COMPT model into the spectral peak energy, E$_{peak}$ as follows:

\begin{equation}
E_{peak} = (2 - \alpha)\times E_{cut}
\end{equation}

in order to be able to compare our results to the previous studies. For all spectral analysis presented here, we used the sum of two blackbody models (BB+BB; \texttt{bbody+bbody on XSPEC}), sum of blackbody and powerlaw model (BB+PO; \texttt{bbody+powerlaw}) and the cutoff-power law model (COMPT; \texttt{cutoffpl}) on XSPEC version 12.9.1 with $\chi^2$ minimization. We generated LB+LB model on XSPEC as given in Equation 2. There are 4 free parameters for the BB+BB (hot and cold blackbody temperatures, and their normalizations), LB+LB (same as BB+BB) and BB+PO (photon index, photon index normalization, temperature and temperature normalization) models. For COMPT model, the number of free parameters is 3 ($\alpha$, $E_{cut}$ and normalization). For PCA and HEXTE joint spectral analysis, we linked PCA and HEXTE parameters (excluding normalizations) equal, which results in 6 free parameters for BB+BB, BB+PO and LB+LB models and 4 free parameters for COMPT model in joint spectral analysis.

\section{Results} \label{sec:specresults}

In general, we find that all of the four models can successfully describe most of the bursts from for all three sources based on the resulting $\chi^2$ statistics. This can be seen in Table 4, in which we present the percentage of spectra that resulted in statistically acceptable fits for each model.  Here, we define the fits to be "acceptable" when the probability of obtaining $\chi^2$  greater than the resulting $\chi^2$ value based on the $\chi^2$ distribution for the corresponding degrees of freedom (DOF), is greater than 0.2. This means that the fits that do not match this criteria have unacceptably large $\chi^2$ values with a low probability of occuring by chance. We see that all of these models can adequately represent the burst spectra at similar levels.

\begin{deluxetable}{cccc}
\tablecaption{Percentage of acceptable spectral fits based on $\chi^2$ probability for the given DOF\label{tab:goodfits}}
\tablewidth{0pt}
\tablehead{
\colhead{Model} & \colhead{SGR J1550$-$5418} & \colhead{SGR 1900+14} &
\colhead{SGR 1806$-$20} \\
}
\startdata
    BB+PO & 73.8 \% & 72.8 \% & 66.0 \% \\
    BB+BB & 61.9 \% & 69.6 \% & 69.2 \% \\
    LB+LB & 71.4 \% & 83.2 \% & 78.7 \% \\
    COMPT & 71.4 \% & 77.6 \% & 67.9 \% \\
\enddata
\end{deluxetable}

In this section, we present the spectral fit results with errors calculated at 1$\sigma$  for all bursts for all models. Note that the comparisons among these spectral models to determine the best-describing ones were done using simulations, which is discussed in detail in the next section. In Tables 4-6 we present the spectral fit results of all four models for each burst, for the three sources separately. For our fits, we fixed the interstellar hydrogen column density to 3.4 $\times$ $10^{22}$ cm$^{-2}$ for SGR J1550$-$5418 (\citealt{halpern2008}), 2.36 $\times 10^{22}$ cm$^{-2}$ for SGR 1900+14 (\citealt{gogus2011}) and 6.8 $\times$ $10^{22}$ cm$^{-2}$ for SGR 1806$-$20 (\citealt{woods2007}; \citealt{gogus2007}). Detailed statistical investigations (i.e. generating reliable parameter and fluence distributions) were possible for SGR 1900+14 and SGR 1806$-$20 due to their large sample size. However, for SGR J1550 5418, the sample size was not sufficiently large to provide reliable distributions for SGR J1550$-$5418 burst spectral parameters and fluence. It is important to note that our burst samples involve partially saturated bursts, for which the reported burst flux should be taken as a lower bound.  

\subsection{SGR 1900+14}\label{subsec:1900}

We list the resulting spectral model parameters and fit statistics for each of SGR 1900+14 bursts in Table 5. In order to generate distributions of spectral fit parameters, we selected the events whose spectral fit parameters yielded less than 50\% uncertainty. We then modeled the distributions with a Gaussian to determine the mean value. We find that distribution of photon indices peak at 0.86 $\pm$ 0.02 with \(\sigma\) = 0.25$\pm$ 0.02 (see the top panel of Figure 4). The distribution of  $E_{cut}$ (Figure 4, top panel) yields a mean value of 14.38 $\pm$ 1.0 keV with a width of \(\sigma\) = 7.96 $\pm$ 1.1 keV. A normal curve fit to the $E_{peak}$ distribution yields a mean of 17.23 $\pm$  1.42 keV for SGR 1900+14 bursts. This is in agreement with \citealt{feroci2004}, in which they conducted their study using BeppoSAX data in the 1.5$-$100 keV band and obtained a mean of 15.8 \(\pm\) 2.3 keV. For the BB+BB model, the mean temperature of the cooler blackbody is 1.76 $\pm$ 0.02 keV (\(\sigma\) = 0.3 $\pm$ 0.02 keV), and the mean temperature of the hotter blackbody is 6.2 $\pm$ 0.2 keV (\(\sigma\) = 4.3 $\pm$ 0.2 keV) (See Figure 4, lower panels). We also computed the 2$-$250 keV flux for all of the bursts, and found that they are between \(4.02\times10^{-9}\) and \(6.9\times10^{-8}\) erg cm$^{-2}$ s$^{-1}$. Finally, we present the fluence distributions for PCA and HEXTE detections in the right panel of Figure 5. We find that the majority of SGR 1900+14 bursts have fluences $<10^{-8}$ erg cm$^{-2}$ with only a few exceptions.

\subsection{SGR 1806$-$20}\label{subsec:1806}

We report all resulting parameters for SGR 1806$-$20 bursts in Table 6. On average, spectral model parameters of SGR 1806$-$20 bursts span narrower intervals compared to those of SGR 1900+14 bursts. We generated spectral parameter distributions for SGR 1806$-$20 with the same procedure as SGR 1900+14. For the COMPT model, we find a photon index distribution mean of 0.62 \(\pm\) 0.005  (\(\sigma\) = 0.22 \(\pm\) 0.005). The exponential cut-off energy distribution peaks at 21.1 \(\pm\) 1.3 keV with \(\sigma\) = 15.58 \(\pm\) 1.5 keV and the inferred $E_{peak}$ mean is 32.02  \(\pm\) 1.84 keV (see Figure 6, top panels). The BB+BB model yields a mean cooler blackbody temperature of 2.02 \(\pm\) 0.02 keV with \(\sigma\) = 0.24 \(\pm\) 0.02 keV. The mean hotter blackbody temperature is 9.6 \(\pm\) 0.2 keV with \(\sigma\) = 2.7 \(\pm\) 0.2 keV (Figure 6, bottom panels). On average, the combined unabsorbed 2$-$250 keV flux of SGR 1806$-$20 bursts are higher than SGR J1550$-$5418 and similar to SGR 1900+14 with a range of \( 4.91\times10^{-9} - 5.46\times10^{-8}\) \(erg/cm^2/s\). Due to the longer average burst duration, burst fluences of SGR 1806$-$20 events tend to be higher than both SGR 1900+14 and SGR J1550$-$5418. We present fluence distributions for PCA and HEXTE detections of SGR 1806$-$20 bursts in the left panel of Figure 5. Please note that for some SGR 1806-20 bursts, an additional normalization term was fixed to to a determined value based on the spectral shape to constrain spectral parameters of some models. For these bursts, the number of free parameters and as a result the degrees of freedom for some models differ (see burst ID : 2 in Table 6 for an example, where BB+PO degrees of freedom differs from BB+BB and LB+LB degrees of freedom by 1 due to a fixed HEXTE normalization term).

\subsection{SGR J1550$-$5418}\label{subsec:1550}
In Table 7, we present spectral fit results of SGR J1550$-$5418 bursts. For this source, we did not construct distribution plots due to the small sample size especially after taking into account the parameter constraint limit. For the COMPT model, the photon index range from $-$0.28 to 1.77 with a mean of 1.21 while the exponential cut-off energy range from 4.30 keV to 118.26 keV, with an average of 54.46 keV. The average \textit{E}$_{peak}$  of SGR J1550$-$5418 bursts calculated using Equation 4 is 44.59 keV, with a minimum of 20.46 keV and a maximum of 77.04 keV, and is consistent with those found by \citealt{lin2012} (39 \(\pm\) 13 keV) and \citealt{vanderhorst2012} (45  \(\pm\) 2.1 keV) using XRT and GBM data for the same source. The combined unabsorbed flux (in the 2$-$250 keV band) varies from \(3.72\times10^{-9}\) to \(2.62\times10^{-8}\) erg cm$^{-2}$ s$^{-1}$. For the BB+BB model, the temperature of the cooler component (in keV) range from 1.02 to 2.6 with a mean of 1.76, and from 5.67 to 29.24 with a mean of 13.71 for the hot blackbody component. Note that the parameter ranges and averages presented here are excluding fits where either one of the upper or lower bound errors are not available so that the parameters constraints are known and possible issues due to local $\chi^2$ minima are excluded.
 
\begin{longrotatetable}
\tabletypesize{\tiny}
\begin{deluxetable*}{c|ccc|ccc|ccc|ccccc}
\tablecaption{Spectral Properties of SGR 1900+14 Bursts. \label{tab:spec1900}}
\tablewidth{0pt}
\tablehead{
Burst & \multicolumn{3}{c|}{BB+BB} & \multicolumn{3}{c|}{BB+PO} & \multicolumn{3}{c}{LB+LB} & \multicolumn{5}{c}{COMPT} \\
ID  & kT$_1$ & kT$_2$ & $\chi^2$ & kT & \(\Gamma\) & $\chi^2$ & kT$_1$ & kT$_2$ & $\chi^2$ &  $E_{cut}$ &  \(\alpha\) & $\chi^2$ & PCA Flux $^{1}$ & HEXTE Flux $^{2}$  \\
   & (keV) & (keV) & /DOF & (keV) &  & /DOF & (keV) & (keV) & /DOF & (keV) & & /DOF & \((erg/cm^2/s)\) & \((erg/cm^2/s)\)  \\
}
\startdata
1 & $2.3 \pm 0.2$ & $24.2^{+7.5}_{-5.8}$&22.0/22 & $2.1^{+0.6}_{-0.4}$ & $1.1 \pm 0.2$ & 18.8/22 & $2.6^{+0.3}_{-0.2}$ & $32.2^{+13.3}_{-8.7}$ & 18.8/22 & $171.2^{-9999}_{-111.8}$ & $1.2 \pm 0.1$ & 25.1/24 & $(2.3\pm 0.1)\times10^{-8}$ & $(9.6^{+3.5}_{-3.4})\times10^{-8}$ \\ 
2 & $1.5 \pm 0.3$ & $5.4^{+1.9}_{-0.9}$&9.54/11 & $1.8^{+10.0}_{-1.8}$ & $1.3^{+0.7}_{-0.6}$ & 10.2/11 & $1.6^{+0.5}_{-0.4}$ & $6.5^{+4.9}_{-1.4}$ & 9.86/11 & $83.1^{-9999}_{-68.8}$ & $1.3^{+0.2}_{-0.5}$ & 10.2/13 & $(6.4^{+0.4}_{-0.5})\times10^{-9}$ & $(2.3^{+2.4}_{-1.4})\times10^{-8}$ \\ 
3 & $1.6 \pm 0.6$ & $11.4^{-9999}_{-6.4}$&9.11/6 & $1.5^{+0.8}_{-1.5}$ & $0.2^{+3.0}_{-0.2}$ & 9.10/6 & $1.7^{+0.8}_{-0.7}$ & $32.4^{-9999}_{-26.4}$ & 9.08/6 & $500.0^{-9999}_{-500.0}$ & $1.1^{+0.3}_{-0.6}$ & 9.95/8 & $(1.2\pm 0.2)\times10^{-9}$ & $(2.1\pm -9999)\times10^{-10}$ \\ 
4 & $1.9 \pm 0.1$ & $6.4^{+1.4}_{-0.9}$&25.6/29 & $2.6 \pm 0.3$ & $1.7 \pm 0.1$ & 25.0/29 & $2.3 \pm 0.1$ & $10.0^{+2.4}_{-2.1}$ & 19.5/29 & $17.7^{+6.1}_{-3.9}$ & $1.1^{+0.1}_{-0.2}$ & 27.5/31 & $(4.7\pm 0.1)\times10^{-8}$ & $(3.8^{+0.6}_{-0.5})\times10^{-8}$ \\ 
5 & $2.1 \pm 0.2$ & $7.5 \pm 1.3$&26.2/25 & $3.7^{+0.3}_{-0.2}$ & $2.7^{+0.4}_{-0.6}$ & 30.7/25 & $2.5^{+0.3}_{-0.2}$ & $8.7^{+1.7}_{-1.5}$ & 23.8/25 & $14.9^{+4.9}_{-3.5}$ & $0.8 \pm 0.2$ & 24.9/27 & $(2.1\pm 0.1)\times10^{-8}$ & $(1.9\pm 0.3)\times10^{-8}$ \\ 
\hline
\enddata
\tablenotetext{1}{PCA Flux Energy Range: 2-30 keV}
\tablenotetext{2}{HEXTE Flux Energy Range: 15-250 keV}
\tablenotetext{3}{All errors are reported at 1 $\sigma$, '$-$9999' indicates error information is not available for the given fit}
\tablenotetext{4}{The number of free parameters are the same for BB+BB, BB+PO and LB+LB models and is 4 for each model and 6 for PCA and HEXTE joint spectral analysis. The number of free parameters is 3 for the COMPT model and is 4 in joint spectral analysis. See Section 3 for a list of free model parameters.}
\tablecomments{Table 5 is available in the machine-readable format in full. The first 5 of 125 data lines are presented here to provide an example of its form.}
\end{deluxetable*}
\end{longrotatetable}

 \begin{figure}[h]\label{fig:1900param} 
  \plottwo{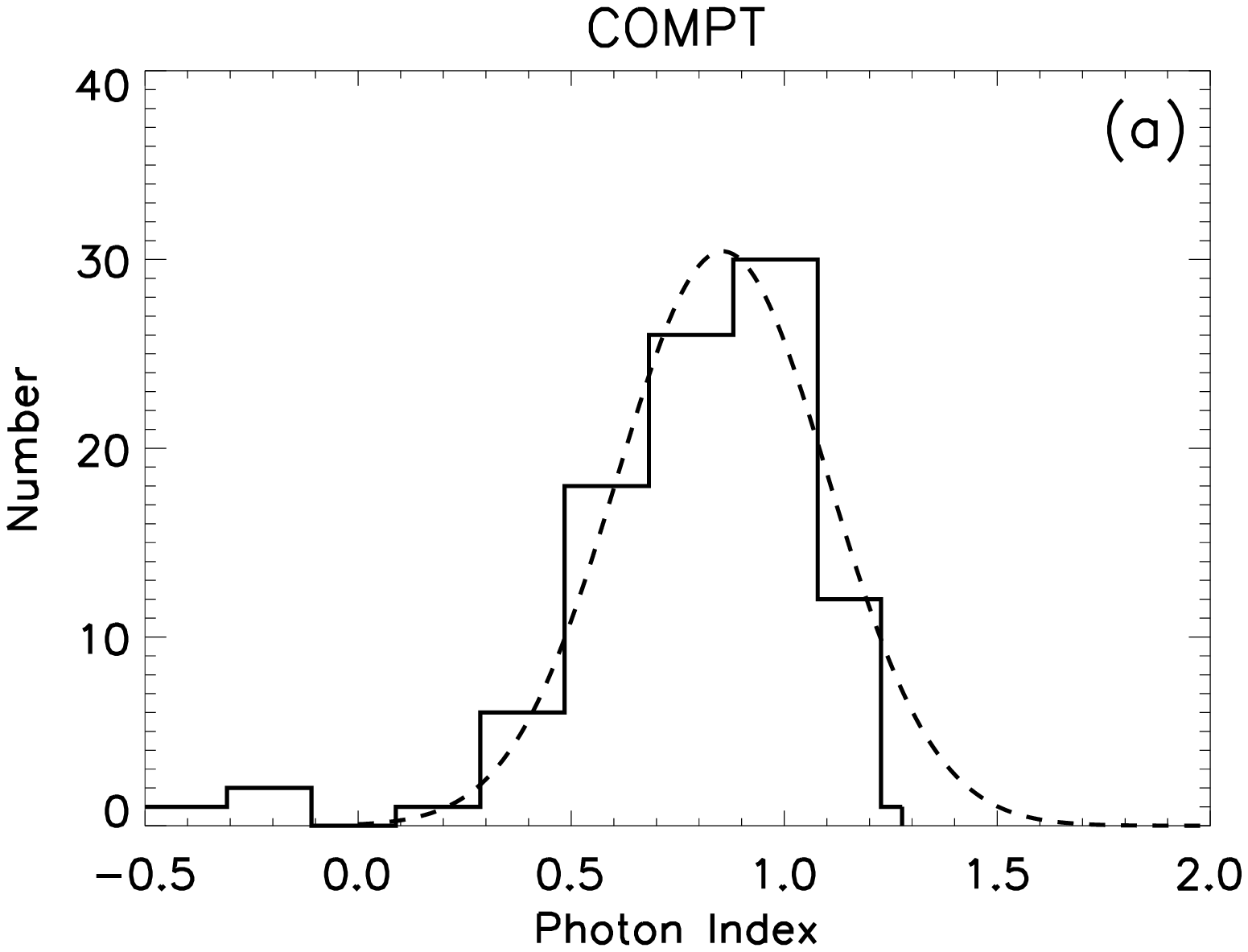}{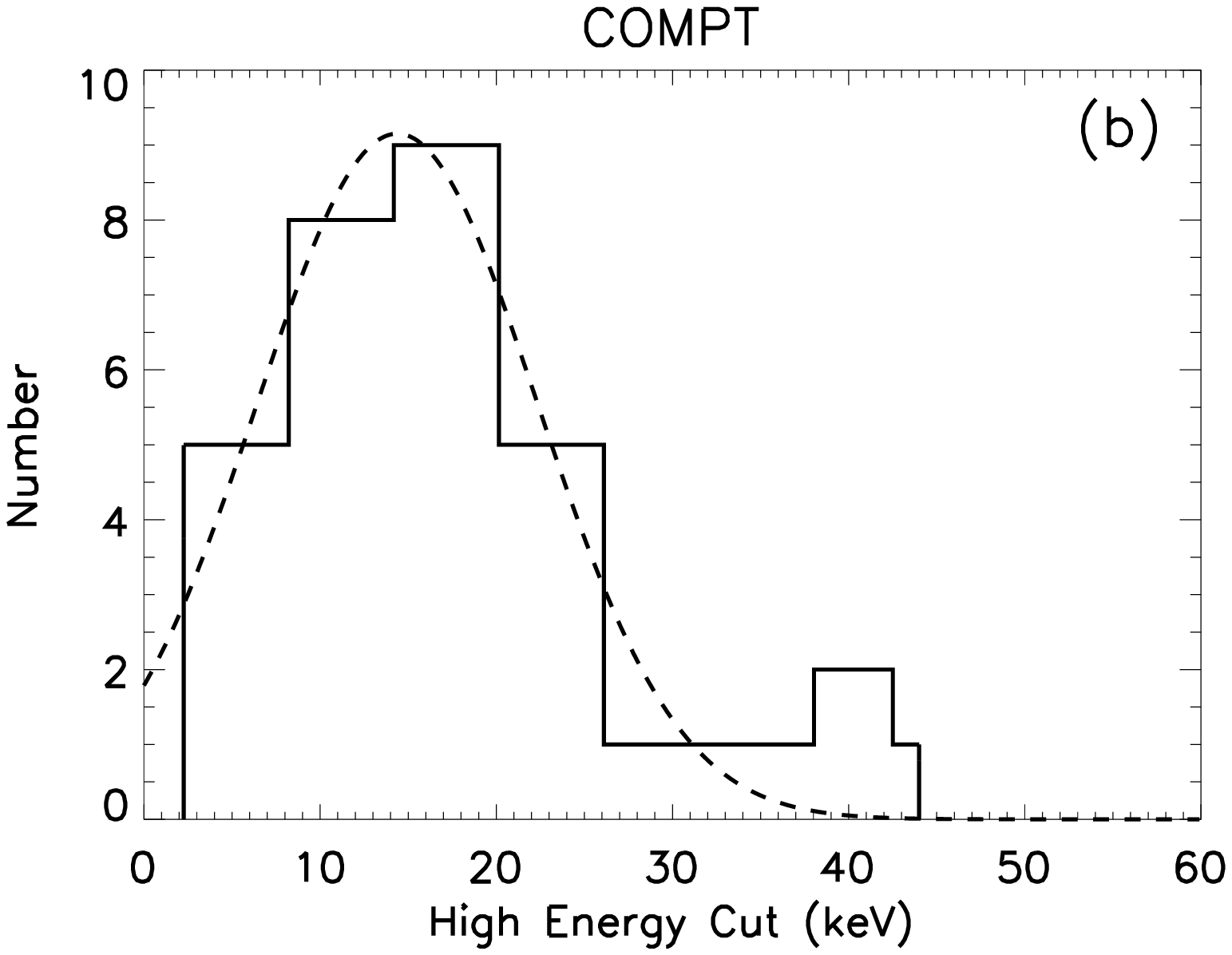}
  \plottwo{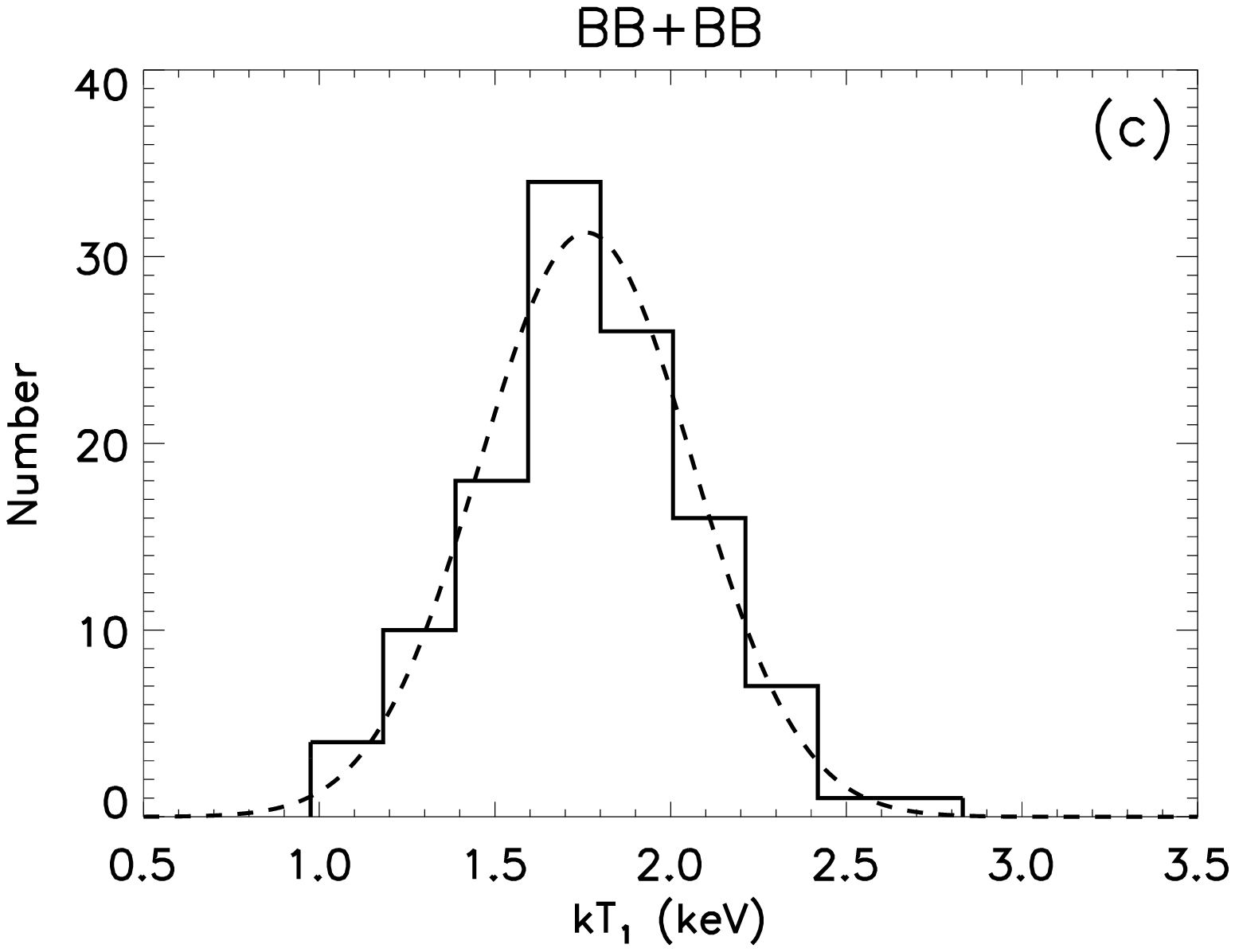}{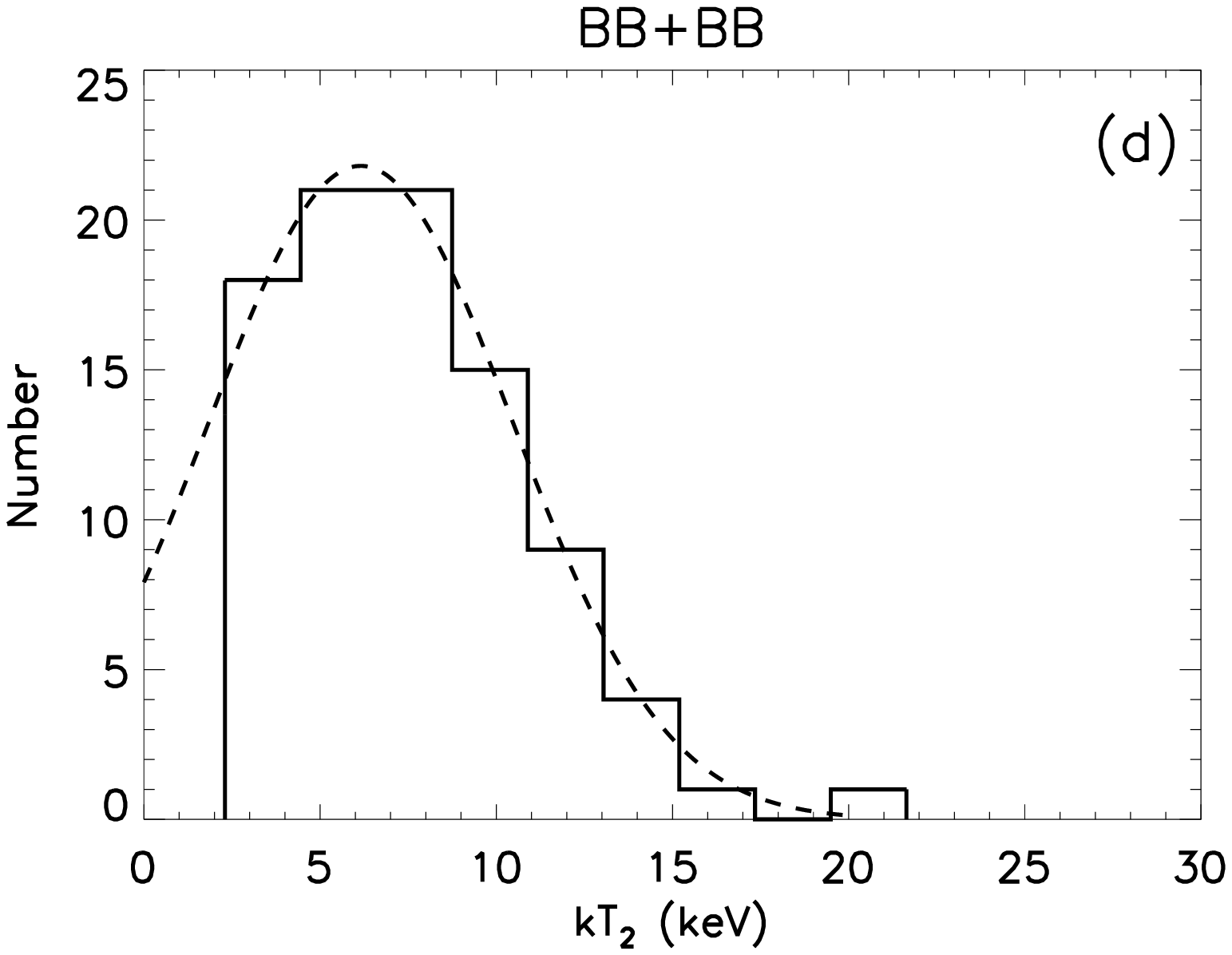} 
 \caption{Distributions of spectral parameters obtained from COMPT and BB+BB model fits to the bursts observed from SGR 1900+14. Top figures are distributions of COMPT model parameters; photon index ($\alpha$) (a) and $E_{cut}$ (b). Bottom figures are distributions  of BB+BB model parameters; first (c) and second (d) blackbody temperatures. Dashed curves represent the Gaussian fits to the parameter distributions.} 
\end{figure}

\begin{longrotatetable}
\tabletypesize{\tiny}
\begin{deluxetable*}{c|ccc|ccc|ccc|ccccc}
\tablecaption{Spectral Properties of SGR 1806$-$20 Bursts. \label{tab:spec1806}}
\tablewidth{0pt}
\tablehead{
Burst & \multicolumn{3}{c|}{BB+BB} & \multicolumn{3}{c|}{BB+PO} & \multicolumn{3}{c}{LB+LB} & \multicolumn{5}{c}{COMPT} \\
ID  & kT$_1$ & kT$_2$ & $\chi^2$ & kT & \(\Gamma\) & $\chi^2$ & kT$_1$ & kT$_2$ & $\chi^2$ &  $E_{cut}$ &  \(\alpha\) & $\chi^2$ & PCA Flux $^{1}$ & HEXTE Flux $^{2}$ \\
   & (keV) & (keV) & /DOF & (keV) &  & /DOF & (keV) & (keV) & /DOF & (keV) & & /DOF & \((erg/cm^2/s)\) & \((erg/cm^2/s\)  \\
}
\startdata
1 & $2.7 \pm 0.2$ & $11.6 \pm 1.6$&54.0/31 & $4.5 \pm 0.3$ & $2.1 \pm 0.1$ & 49.0/32 & $3.3 \pm 0.3$ & $14.1^{+2.0}_{-1.9}$ & 47.5/31 & $30.7^{+8.0}_{-6.3}$ & $0.9 \pm 0.1$ & 49.4/33 & $(3.1\pm 0.1)\times10^{-8}$ & $(4.1\pm 0.5)\times10^{-8}$ \\ 
2 & $2.6^{+0.5}_{-0.4}$ & $9.6^{+1.5}_{-1.3}$&25.2/16 & $8.2^{+1.5}_{-2.5}$ & $1.2^{+0.9}_{-0.2}$ & 25.2/15 & $3.0^{+0.7}_{-0.8}$ & $10.7^{+2.8}_{-1.7}$ & 24.8/16 & $22.3^{+9.0}_{-5.9}$ & $0.5 \pm 0.2$ & 24.6/17 & $(2.1\pm 0.1)\times10^{-8}$ & $(3.2\pm 0.6)\times10^{-8}$ \\ 
3 & $2.6 \pm 0.3$ & $14.9^{+4.3}_{-3.5}$&7.68/13 & $3.9^{+1.1}_{-0.7}$ & $1.5 \pm 0.3$ & 12.4/14 & $3.0 \pm 0.4$ & $17.4^{+5.4}_{-3.8}$ & 8.36/14 & $69.0^{+97.5}_{-32.4}$ & $1.1 \pm 0.2$ & 12.3/15 & $(1.2\pm 0.1)\times10^{-8}$ & $(1.8^{+0.7}_{-0.5})\times10^{-8}$ \\ 
4 & $3.1^{+0.7}_{-1.0}$ & $23.1^{-9999}_{-17.7}$&0.35/3 & $3.9^{+2.0}_{-0.5}$ & $1.6^{+2.2}_{-0.8}$ & 0.36/3 & $4.1^{-9999}_{-4.1}$ & $10.5^{+43.3}_{-7.1}$ & 0.48/3 & $16.1 \pm {-9999} $ & $0.5 \pm 0.3$ & 0.88/5 & $(1.7^{+0.2}_{-0.4})\times10^{-9}$ & $(2.7^{+3.3}_{-1.4})\times10^{-9}$ \\ 
5 & $2.7 \pm 0.1$ & $10.0 \pm 0.6$&55.7/63 & $8.8 \pm 0.5$ & $1.0^{+0.2}_{-0.1}$ & 81.5/63 & $3.2 \pm 0.3$ & $11.5 \pm 0.8$ & 53.9/63 & $21.9^{+2.1}_{-1.9}$ & $0.4 \pm 0.1$ & 57.9/65 & $(9.1\pm 0.2)\times10^{-9}$ & $(1.2\pm 0.1)\times10^{-8}$ \\ 
\hline
\enddata
\tablenotetext{1}{PCA Flux Energy Range: 2-30 keV}
\tablenotetext{2}{HEXTE Flux Energy Range: 15-250 keV}
\tablenotetext{3}{All errors are reported at 1 $\sigma$, $-9999$ indicates error information is not available for the given fit}
\tablenotetext{4}{The number of free parameters are the same for BB+BB, BB+PO and LB+LB models and is 4 for each model and 6 for PCA and HEXTE joint spectral analysis. The number of free parameters is 3 for the COMPT model and is 4 in joint spectral analysis. See Section 3 for a list of free model parameters.}
\tablenotetext{5}{Note that for a some bursts, one normalization term was fixed to constrain spectral parameters and therefore the number of free parameters and the corresponding degrees of freedom may differ.}
\tablecomments{Table 6 is available in the machine-readable format in full. The first 5 of 221 data lines are presented here to provide an example of its form.}
\end{deluxetable*}
\end{longrotatetable}

\begin{figure}[h] \label{fig:fluence} 
\begin{centering}
\plottwo{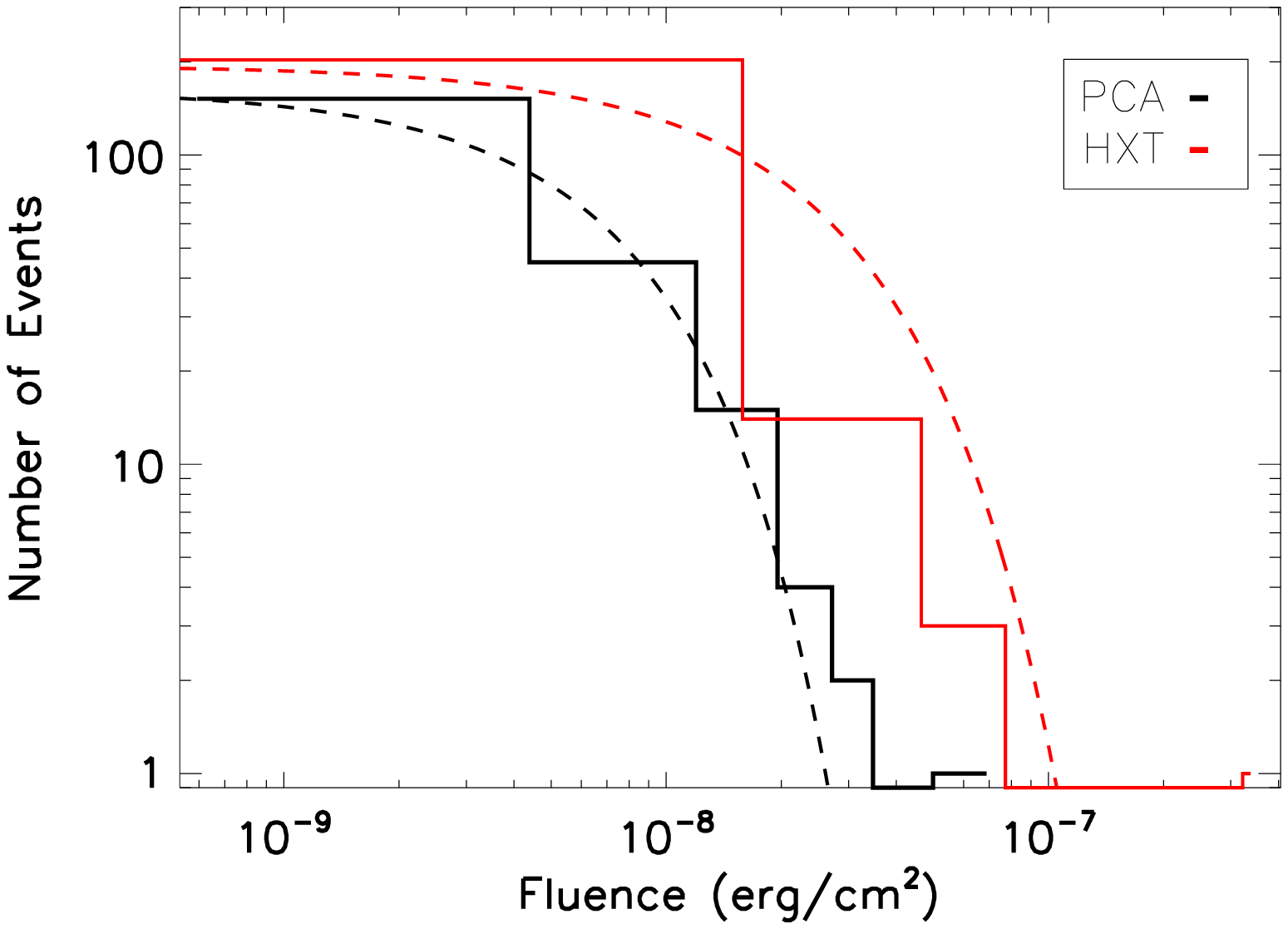}{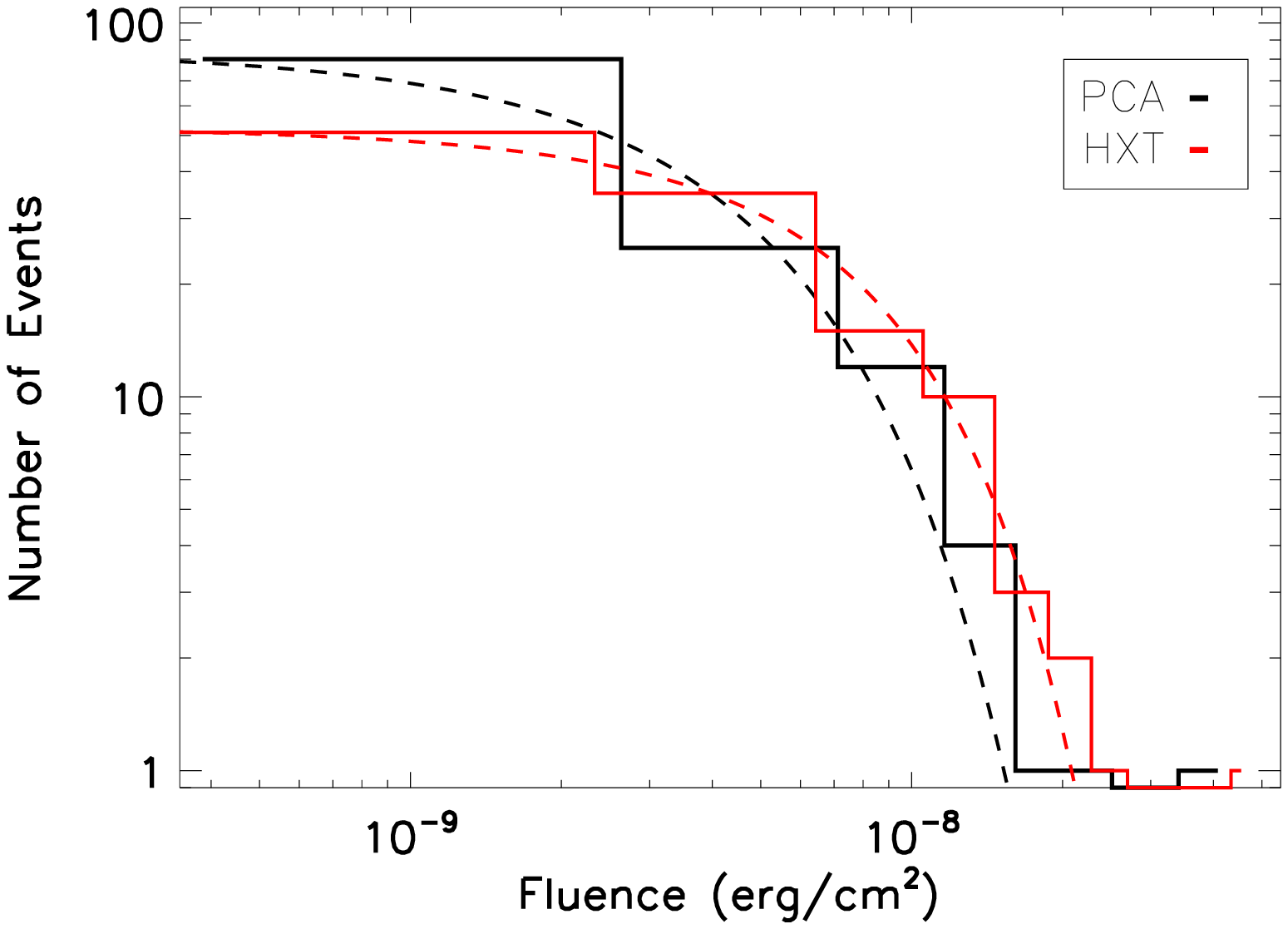} 
\caption{Fluence distributions of SGR 1806$-$20 (left) and SGR 1900+14 (right) bursts. Fits are calculated with the COMPT model with PCA fluence shown in black and HEXTE in red. Dashed lines represent Gaussian fits to the fluence distributions.} 
\end{centering}
\end{figure}

\begin{figure}[ht] \label{fig:1806param} 
    \plottwo{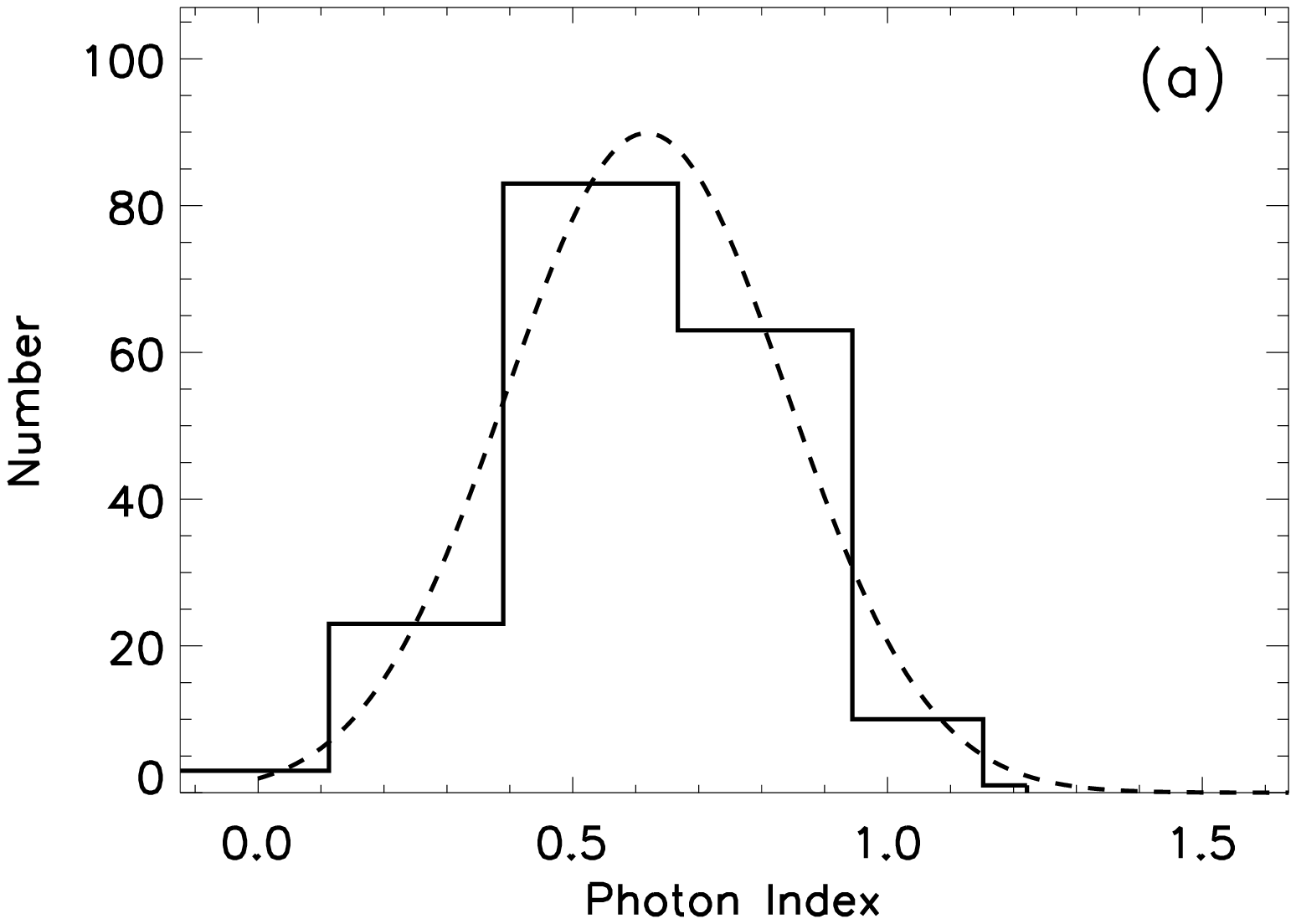}{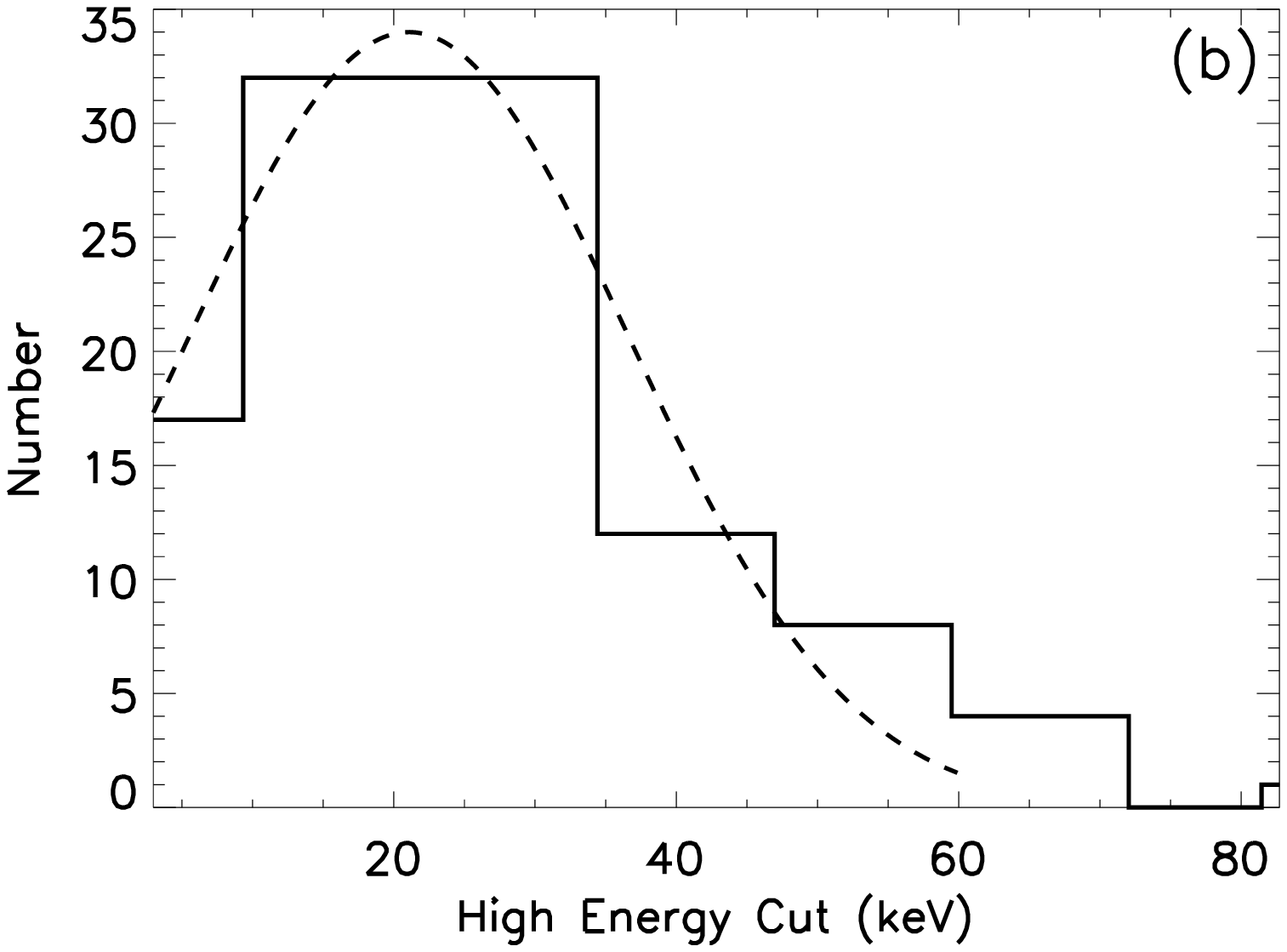}
    \plottwo{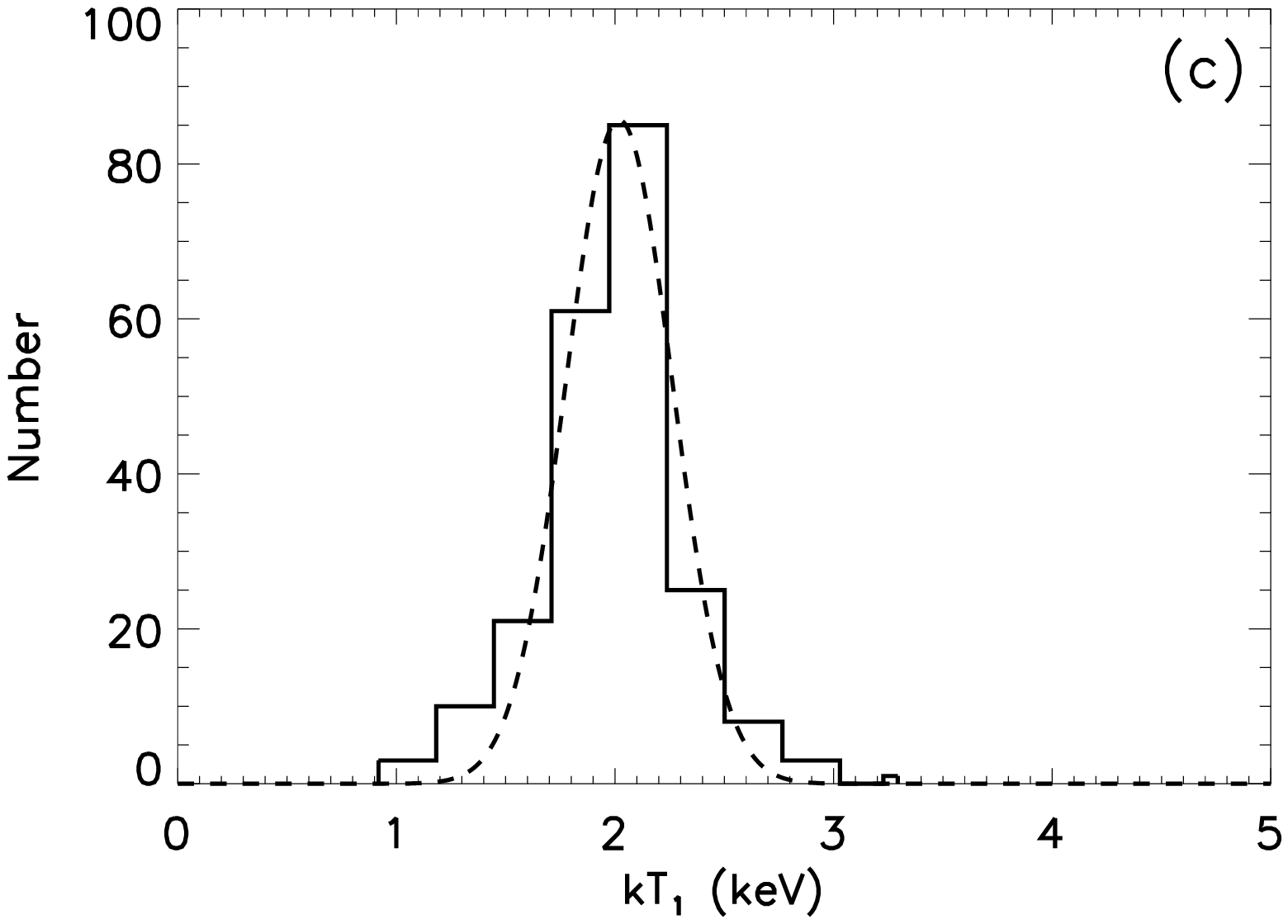}{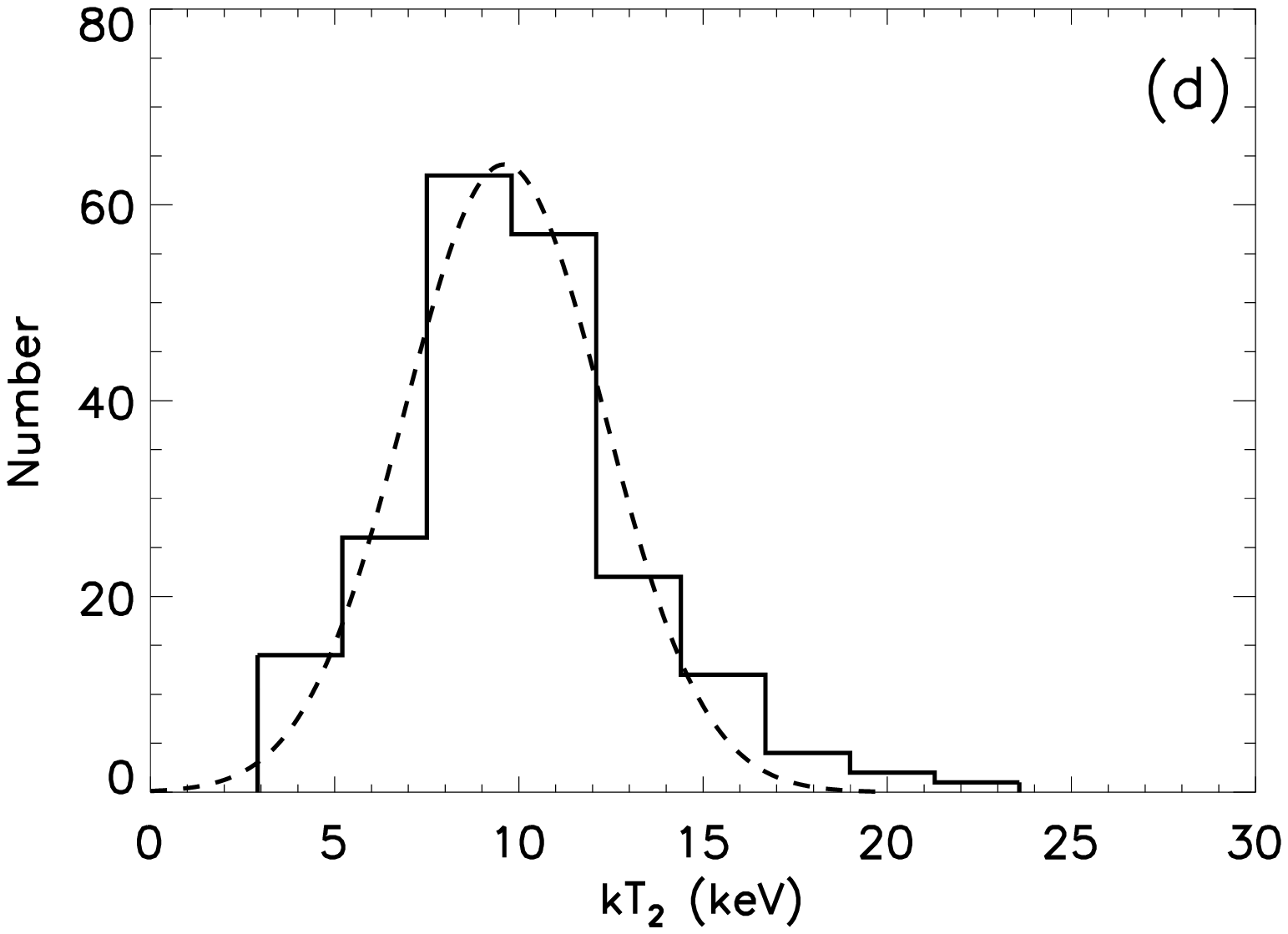} 
 \caption{Distributions of spectral parameters obtained from COMPT and BB+BB model fits to the bursts observed from SGR 1806$-$20. Top figures are distributions of COMPT model parameters; photon index (a) and $E_{cut}$ (b). Bottom figures are distributions  of BB+BB model parameters; first (c) and second (d) blackbody temperatures. Dashed curves represent the Gaussian fits to the parameter distributions.} 
\end{figure}

 \begin{longrotatetable}
\tabletypesize{\tiny}
\begin{deluxetable*}{c|ccc|ccc|ccc|ccccc}
\tablecaption{Spectral Properties of SGR J1550$-$5418 Bursts. \label{tab:spec1550}}
\tablewidth{0pt}
\tablehead{
Burst & \multicolumn{3}{c|}{BB+BB} & \multicolumn{3}{c|}{BB+PO} & \multicolumn{3}{c}{LB+LB} & \multicolumn{5}{c}{COMPT} \\
ID  & kT$_1$ & kT$_2$ & $\chi^2$ & kT & \(\Gamma\) & $\chi^2$ & kT$_1$ & kT$_2$ & $\chi^2$ &  $E_{cut}$ &  \(\alpha\) & $\chi^2$ & PCA Flux $^{1}$ & HEXTE Flux $^{2}$  \\
   & (keV) & (keV) & /DOF & (keV) &  & /DOF & (keV) & (keV) & /DOF & (keV) & & /DOF & \((erg/cm^2/s)\) & \((erg/cm^2/s)\)  \\
}
\startdata
1 & $2.0^{+0.4}_{-0.3}$ & $9.2^{-9999}_{-3.1}$&3.21/3 & $2.1^{+0.7}_{-0.3}$ & $0.6^{+1.1}_{-2.0}$ & 3.23/3 & $2.5 \pm 0.5$ & $27.6^{-9999}_{-20.6}$ & 3.31/3 & $10.1^{+50.6}_{-5.5}$ & $0.6^{+0.8}_{-1.0}$ & 5.64/5 & $(1.3\pm 0.2)\times10^{-8}$ & $(3.7^{+4.5}_{-1.4})\times10^{-9}$ \\ 
2 & $2.2 \pm 0.2$ & $29.2^{+12.8}_{-8.9}$&12.1/8 & $2.1^{+0.3}_{-0.2}$ & $1.1^{+0.7}_{-0.6}$ & 13.9/8 & $2.6^{+0.4}_{-0.3}$ & $36.2^{+23.1}_{-12.9}$ & 12.3/8 & $4.3^{+2.3}_{-1.3}$ & $-0.3^{+0.7}_{-0.8}$ & 17.9/10 & $(6.2\pm 0.7)\times10^{-9}$ & $(2.1^{+0.6}_{-0.5})\times10^{-9}$ \\ 
3 & $1.6 \pm 0.2$ & $7.6^{+1.4}_{-1.2}$&6.22/8 & $7.6^{+1.1}_{-1.5}$ & $1.5 \pm 0.1$ & 9.26/8 & $1.7 \pm 0.3$ & $8.6^{+1.7}_{-1.4}$ & 6.57/8 & $25.6^{+13.0}_{-7.9}$ & $1.2 \pm 0.2$ & 10.1/10 & $(2.0\pm 0.2)\times10^{-8}$ & $(2.2^{+0.5}_{-0.4})\times10^{-8}$ \\ 
4 & $2.1 \pm 0.2$ & $20.7^{+4.5}_{-3.4}$&17.4/14 & $2.0 \pm 0.3$ & $1.2 \pm 0.2$ & 12.6/14 & $2.5 \pm 0.3$ & $27.7^{+7.6}_{-5.5}$ & 13.8/14 & $500.0^{-9999}_{-212.9}$ & $1.5 \pm 0.1$ & 21.5/16 & $(1.4\pm 0.1)\times10^{-8}$ & $(3.4\pm -9999)\times10^{-8}$ \\ 
5 & $1.7 \pm 0.2$ & $20.1^{+2.6}_{-2.3}$&20.6/17 & $0.6^{+0.5}_{-0.4}$ & $1.2 \pm 0.1$ & 22.4/17 & $1.7 \pm 0.3$ & $25.1^{+3.8}_{-3.3}$ & 16.8/17 & $500.0^{-9999}_{-327.1}$ & $1.3^{+0.1}_{-0.2}$ & 24.4/19 & $(1.4\pm 0.1)\times10^{-8}$ & $(4.6^{+0.4}_{-0.8})\times10^{-8}$ \\ 
\hline
\enddata
\tablenotetext{1}{PCA Flux Energy Range: 2-30 keV}
\tablenotetext{2}{HEXTE Flux Energy Range: 15-250 keV}
\tablenotetext{3}{All errors are reported at 1 $\sigma$, $-$9999 indicates error information is not available for the given fit}
\tablenotetext{4}{The number of free parameters are the same for BB+BB, BB+PO and LB+LB models and is 4 for each model and 6 for PCA and HEXTE joint spectral analysis. The number of free parameters is 3 for the COMPT model and is 4 in joint spectral analysis. See Section 3 for a list of free model parameters.}
\tablecomments{Table 7 is available in the machine-readable format in full. The first 5 of 42 data lines are presented here to provide an example of its form.}
\end{deluxetable*}
\end{longrotatetable}

\subsection{Companion Web-catalog of Magnetar Burst Spectral and Temporal Characteristics}

We also made the results of our analyses available at a companion web-catalog which includes general properties (Burst time, total photon counts, peak counts and 5 PCU plots where the amount of time excluded due to saturation for saturated bursts can also be found), temporal analysis results (including burst duration with start and end times obtained with Bayesian Blocks algorithm for single peak and multi-peak bursts) and spectral analysis results (BB+BB temperatures, BB+PO temperature and photon index, COMPT photon index and cut-off energy and LB+LB temperatures for single-peak bursts). Note that our spectral analysis involves single-peak bursts only. Additionally, the web-catalog provides all temporal and spectral data in FITS format for interested researchers to download and perform their independent investigations. The address of the companion web-catalog is http://magnetars.sabanciuniv.edu  

\section{Simulations for Model Comparisons}\label{sec:simu}

Even though one of the four models that were employed to fit the broadband X-ray spectra of magnetar bursts yields the minimum reduced $\chi^2$ value, it is statistically not possible to disregard the alternatives simply by a \({\Delta}{\chi}^2\) test. This issue becomes more complicated given the fact that the COMPT model involves one less free parameter than thermal models, resulting in, on average, one more degrees of freedom. The additional degree of freedom enhances the fitting power of COMPT in the cases where the spectrum could be statistically represented by two or more models. In such cases, competing models can be better compared by simulations based on fit results. 

To achieve this objective, we performed extensive simulations for each burst as follows: Overall the COMPT model is expected to perform the best in fitting magnetar burst spectra based on number of parameters. Therefore, we took the COMPT model as the null hypothesis and generated 1000 spectra using the resulting COMPT fit parameters for each burst whose COMPT model parameters were sufficiently constrained (parameter error to be less than 50\% of the parameter). As an alternative hypothesis (test model), we selected one of the three thermal models, whose reduced \({\chi}^2\) value was the smallest. Note that the remaining models have equal number of parameters and a simple  \({\Delta}{\chi}^2\) test is applicable for comparison.

When the test model did not provide well-constrained parameters (i.e. less than 50\% errors), we have selected the next model with the least reduced \({\chi}^2\) value to be the test model. If none of the test models provided well-constrained parameters, we did not perform the simulation for that event. We found that four out of 42 events examined for SGR J1550$-$5418, 21 out of 125 events examined for SGR 1900+14 and 77 out of 221 bursts from SGR 1806$-$20 provided such well-constrained parameters for the seed and test models, and were included in our simulations. We have then fit the 1000 generated spectra for each burst with the COMPT model and its alternative test model.

We used a significance level (i.e. probability of rejecting the null hypothesis given that it is true) of 0.05 for each burst included in the simulation. For a \({\chi}^2\) distribution with dof = 1 (since the degrees of freedom on the test and seed model differ by one in each case), this corresponds to a \({\chi}^2\) value of 3.84. Therefore, we defined our rejection region of the null hypothesis (i.e. when we accept the test model) as the region where the test model \({\chi}^2\) is less than the seed model \({\chi}^2\) by at least 3.84. We suggest that if a truly Comptonized spectrum in fact provides better fit statistics within 0.05 significance, then our fit results where COMPT provides lower \({\chi}^2\) values indicates the true emission mechanism most likely is Comptonized rather than thermal at least on time-averaged spectra. 

Similar to the procedure in \citealt{lin2012}, we define our p-value to be the fraction of simulated spectra better fitted by COMPT model with the 0.05 significance. If the p-value exceeds 0.9, we conclude that the COMPT model provides better fit statistics than the test model when it is the underlying emission mechanism. In Figure 7, we present the distributions for the difference between seed model (COMPT) \({\chi}^2\) and test model \({\chi}^2\) for three example events of SGR 1806$-$20 with different test models for a visual description of simulation results. Our rejection region of COMPT model (null hypothesis) is where $\chi^2_{COMPT} - \chi^2_{Test Model}$ $\geq$ 3.84, as described above. We also list the resulting p-values for the entire sample in Table 8.  

\begin{figure}\label{fig:simuchi}
\includegraphics[width= 0.45\textwidth]{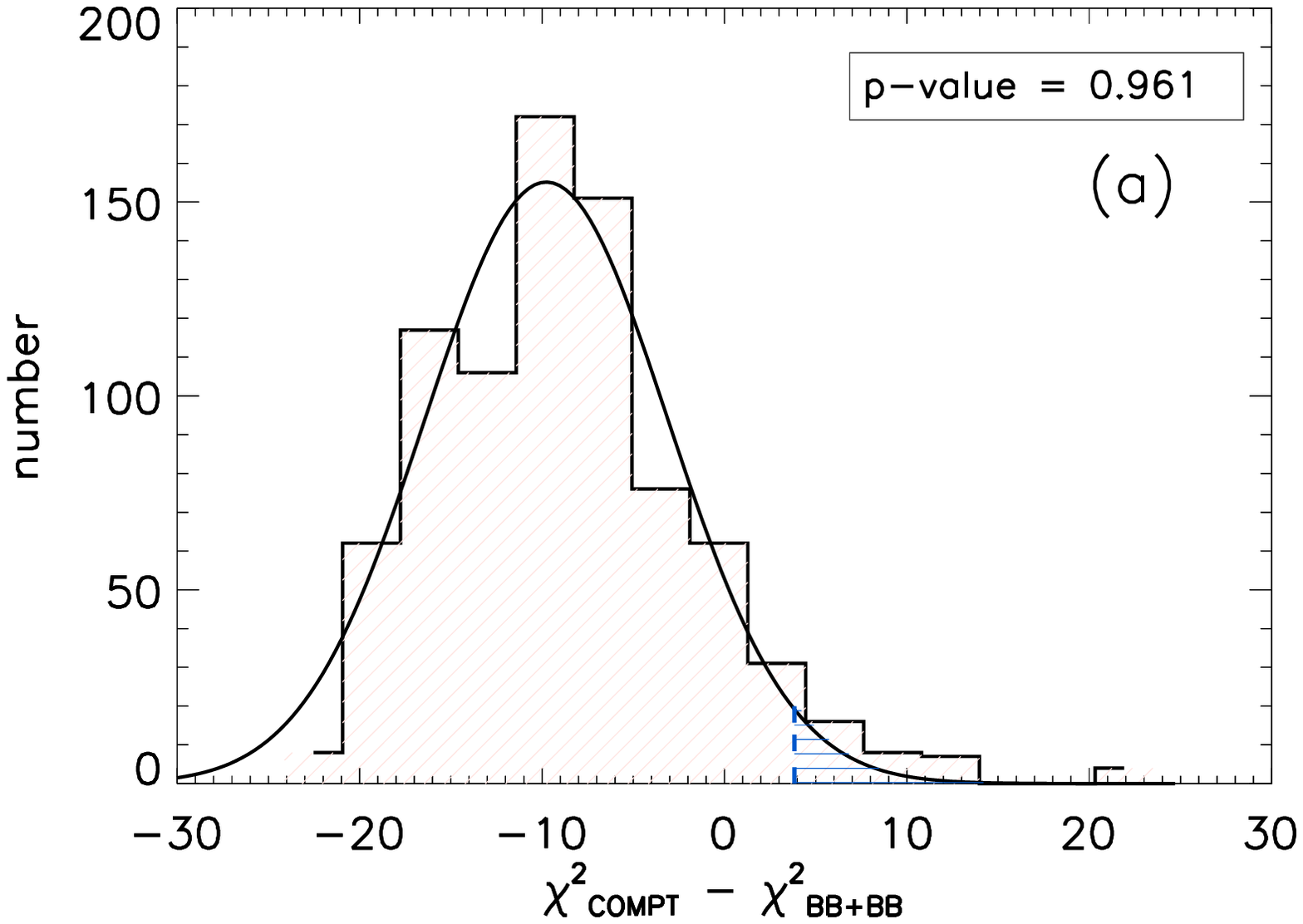}
\includegraphics[width= 0.45\textwidth]{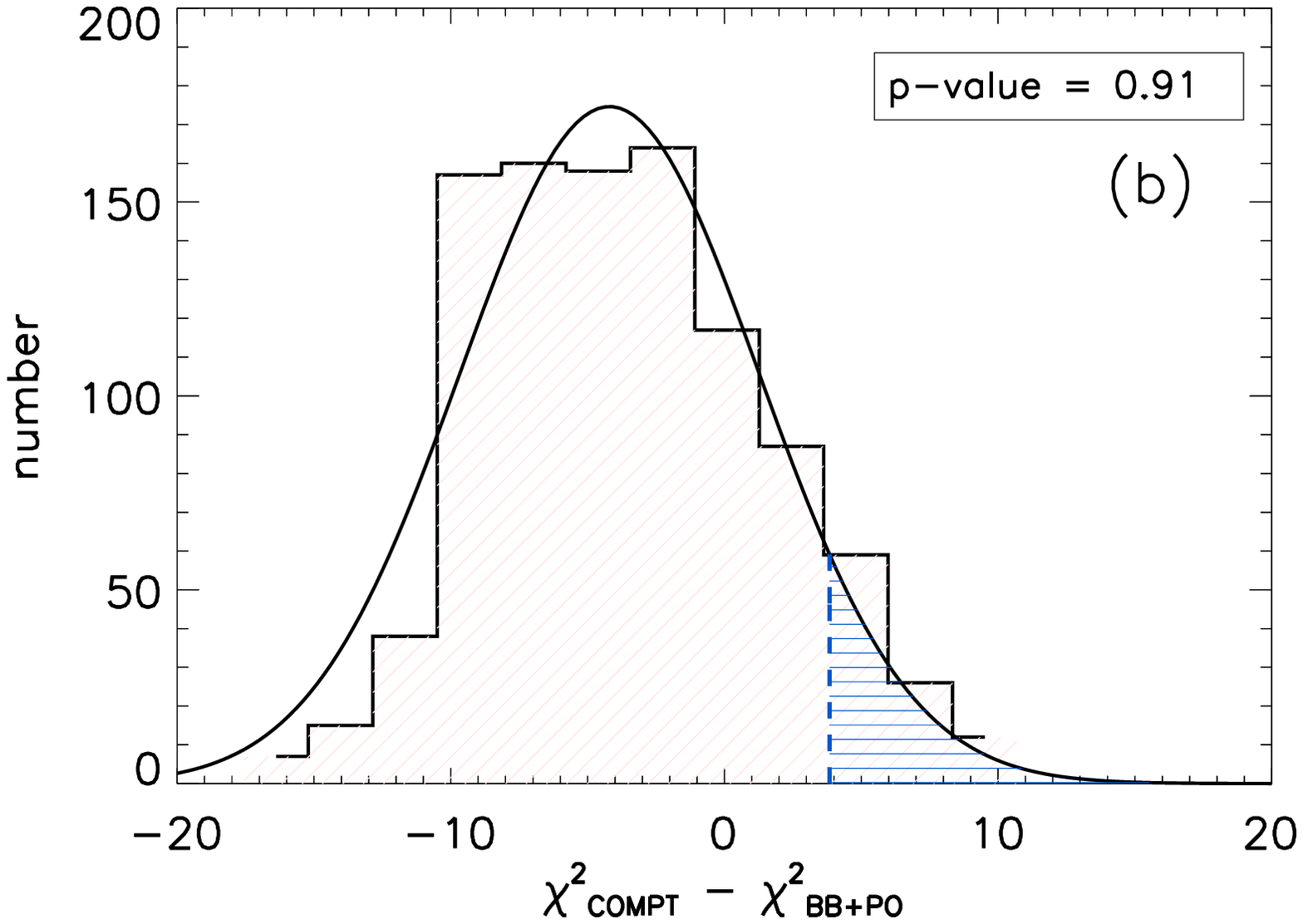} \\
\centering
\includegraphics[width= 0.45\textwidth]{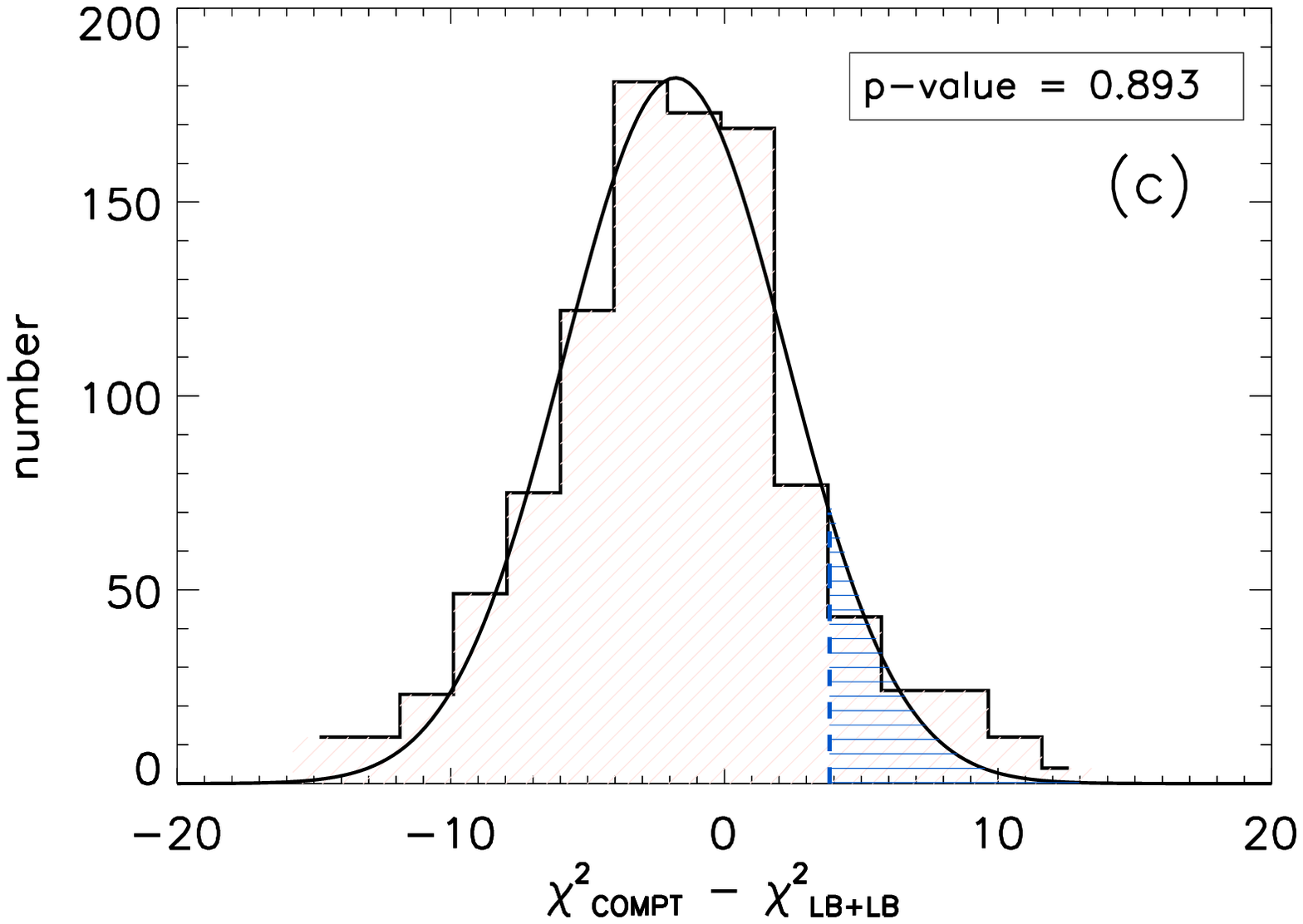}
\caption{COMPT $\chi^2$ - Test Model $\chi^2$ distributions for three SGR 1806$-$20 bursts. Blue shaded regions represent rejection regions of the seed (COMPT) model (where $\chi^2_{COMPT} - \chi^2_{Test Model} > 3.84$) (a) Test Model : BB+BB for Burst ID: 20 (b) Test Model: BB+PO for Burst ID: 1 (c) Test Model: LB+LB for Burst ID: 5}
\end{figure}

We found that the COMPT model is the most frequently preferred model based on simulation results: The COMPT model provides significantly better fit statistics in more than 90\% of trials for 16/19 events compared to the BB+PO model, 12/17 events compared to the BB+BB model and 41/66 events compared to the LB+LB model. Overall, COMPT provides statistically better fits to more than 67.6$\%$ of cases against its alternatives.

Also, LB+LB model emerges as the best fitting model within thermal models, providing better fits than BB+BB and BB+PO models for the majority (66 out of 102) of events. It is important to reinstate that LB+LB was selected as the test model in each of these cases because it  provided the lowest reduced ${\chi}^2$ value among the thermal models with well-constrained parameters and that it is possible to compare the thermal models with a simple $\Delta {\chi}^2$ test since these models have the same number of parameters. 

To check whether the simulation procedure forms a bias towards COMPT, we repeated the same procedure with BB+BB as the null hypothesis (seed model) for one event with COMPT as the alternative hypothesis (test model). In this reverse simulation scenario, we similarly defined our rejection region for the null hypothesis as when COMPT ${\chi}^2$ value was less than BB+BB ${\chi}^2$ by at least ${{\chi}^2}_{0.05,1}$ = 3.84. BB+BB model was accepted in 100\% of trials when it acted as the seed model. COMPT model was accepted in 99.7\% of trials when COMPT was the seed model in the original simulation for the same event. By comparing these results, we concluded that the simulation procedure accepts the inherent emission mechanism within the level of significance with no bias towards any model. Therefore, we continued the simulations with COMPT as the seed model.

\begin{deluxetable*}{ccc|ccc|ccc|ccc}\label{tab:pval}
\tabletypesize{\scriptsize}
\tablecaption{P-values of Simulated Bursts \label{tab:pvals}}
\tablewidth{0pt}
\tablehead{
\multicolumn{3}{c|}{SGR J1550$-$5418} & \multicolumn{3}{c|}{SGR 1900+14} & \multicolumn{3}{c}{SGR 1806$-$20} & \multicolumn{3}{c}{SGR 1806$-$20-continued}\\
Test & p-value & Burst ID & Test & p-value & Burst ID & Test & p-value & Burst ID  & Test & p-value & Burst ID\\
Model & & & Model & & & Model & & & Model & &
}
\startdata
BB+PO & 0.961 & 24 &	BB+BB & 0.930 & 85 &	BB+BB & 0.961 & 20 &	LB+LB & 0.879 & 46 \\
BB+PO & 0.984 & 28 &	BB+BB & 0.997 & 101 &	BB+BB & 0.848 & 31 &	LB+LB & 0.910 & 47 \\
LB+LB & 1.000 & 26 &	BB+BB & 0.921 & 102 &	BB+BB & 0.870 & 44 &	LB+LB & 0.941 & 49 \\
LB+LB & 1.000 & 36 &	BB+BB & 0.930 & 124 &	BB+BB & 0.996 & 69 &	LB+LB & 1.000 & 50 \\
 & & &	BB+PO & 0.895 & 112 &	BB+BB & 0.897 & 70 &	LB+LB & 0.915 & 52 \\
 & & &	LB+LB & 0.946 & 4 &	BB+BB & 0.935 & 82 &	LB+LB & 0.926 & 66 \\
 & & &	LB+LB & 0.865 & 5 &	BB+BB & 0.887 & 93 &	LB+LB & 0.828 & 67 \\
 & & &	LB+LB & 0.939 & 81 &	BB+BB & 0.948 & 152 &	LB+LB & 0.943 & 68 \\
 & & &	LB+LB & 0.870 & 84 &	BB+BB & 0.978 & 153 &	LB+LB & 1.000 & 74 \\
 & & &	LB+LB & 0.942 & 87 &	BB+BB & 0.881 & 170 &	LB+LB & 0.864 & 78 \\
 & & &	LB+LB & 0.952 & 88 &	BB+BB & 0.929 & 185 &	LB+LB & 0.980 & 79 \\
 & & &	LB+LB & 0.863 & 99 &	BB+BB & 0.945 & 186 &	LB+LB & 0.928 & 80 \\
 & & &	LB+LB & 0.837 & 103 &	BB+BB & 0.985 & 188 &	LB+LB & 0.880 & 81 \\
 & & &	LB+LB & 0.919 & 105 &	BB+PO & 0.910 & 1 &	LB+LB & 0.901 & 89 \\
 & & &	LB+LB & 0.927 & 106 &	BB+PO & 0.940 & 6 &	LB+LB & 0.909 & 95 \\
 & & &	LB+LB & 0.892 & 108 &	BB+PO & 0.919 & 28 &	LB+LB & 0.919 & 99 \\
 & & &	LB+LB & 0.938 & 114 &	BB+PO & 0.935 & 35 &	LB+LB & 0.864 & 108 \\
 & & &	LB+LB & 0.926 & 115 &	BB+PO & 0.902 & 37 &	LB+LB & 0.873 & 110 \\
 & & &	LB+LB & 0.996 & 116 &	BB+PO & 0.893 & 55 &	LB+LB & 0.811 & 129 \\
 & & &	LB+LB & 0.955 & 122 &	BB+PO & 0.965 & 56 &	LB+LB & 0.992 & 131 \\
 & & &	LB+LB & 0.915 & 123 &	BB+PO & 0.922 & 83 &	LB+LB & 0.860 & 132 \\
 & & &	 & & &	BB+PO & 0.910 & 92 &	LB+LB & 0.934 & 134 \\
 & & &	 & & &	BB+PO & 0.911 & 98 &	LB+LB & 0.944 & 135 \\
 & & &	 & & &	BB+PO & 0.898 & 102 &	LB+LB & 1.000 & 136 \\
 & & &	 & & &	BB+PO & 0.996 & 103 &	LB+LB & 0.988 & 138 \\
 & & &	 & & &	BB+PO & 0.934 & 105 &	LB+LB & 0.926 & 140 \\
 & & &	 & & &	BB+PO & 1.000 & 133 &	LB+LB & 0.969 & 141 \\
 & & &	 & & &	BB+PO & 0.988 & 149 &	LB+LB & 0.875 & 142 \\
 & & &	 & & &	BB+PO & 0.932 & 168 &	LB+LB & 0.983 & 148 \\
 & & &	 & & &	LB+LB & 0.893 & 5 &	LB+LB & 0.916 & 155 \\
 & & &	 & & &	LB+LB & 0.821 & 7 &	LB+LB & 0.866 & 158 \\
 & & &	 & & &	LB+LB & 0.883 & 10 &	LB+LB & 0.942 & 164 \\
 & & &	 & & &	LB+LB & 0.842 & 22 &	LB+LB & 1.000 & 169 \\
 & & &	 & & &	LB+LB & 0.966 & 23 &	LB+LB & 0.891 & 172 \\
 & & &	 & & &	LB+LB & 0.840 & 26 &	LB+LB & 0.978 & 177 \\
 & & &	 & & &	LB+LB & 0.846 & 30 &	LB+LB & 0.875 & 182 \\
 & & &	 & & &	LB+LB & 0.848 & 34 &	LB+LB & 0.921 & 183 \\
 & & &	 & & &	LB+LB & 0.911 & 39 &	LB+LB & 0.867 & 184 \\
 & & &	 & & &	LB+LB & 0.949 & 41 &	& & \\
\enddata
\end{deluxetable*}

\section{Discussion}\label{sec:disc}

The models involving non-thermal and thermal emission processes are commonly discussed in the context of magnetar bursts (e.g. \citealt{lin2012}; \citealt{vanderhorst2012}; \citealt{feroci2004}). In the Comptonization viewpoint, the photons emerging from the ignition point are repeatedly upscattered by the e$^{\pm}$-pairs present in the corona. The density and optical thickness of the corona, the incoming photon distribution and the electron temperature set a spectral break point for energy, realized as the peak energy parameter for the power law shaped Comptonized spectrum. For magnetar bursts, the Comptonized spectrum resembles the models for accretion disks and AGN, but the underlying mechanism differs due to the strong magnetic field of magnetars. The corona of hot electrons may emerge in the inner dynamic magnetosphere due to field line twisting as discussed by \citealt{thompsonlyutikovkulkarni2002}, \citealt{thompsonbeloborodov2005}, \citealt{beloborodovthompson2007a} and \citealt{nobili2008}. The magnetospheric corona could give rise to a similar Comptonization process and therefore upscatter emergent photons. This type of corona are expected to be anisotropic due to the intense and likely multipolar magnetic field around the magnetar. The anisotropy of the corona sets a different slope for the emission spectrum. The emission spectrum however is similar in its exponential tail and peak energy, which is now controlled by the thermal and spatial parameters of the corona in the magnetosphere. The distinction between persistent and burst emissions are further explained in this model as the bursts may be triggered closer to the surface where the density of e$^{\pm}$-pairs is high, and persistent $<$ 10 keV signals may originate at higher altitudes with a lower e$^{\pm}$ density (\citealt{beloborodovthompson2007a}).

The alternative approach to interpret magnetar burst spectra is the thermal emission due to a short-lasting plasma of electron-photon pairs in quasi-thermal equilibrium, usually described with the superposition of two blackbody functions (see e.g., \citealt{feroci2004}; \citealt{vanderhorst2012}). This dual blackbody scheme approximates a continuum temperature gradient due to the total energy dissipation of photons throughout the magnetosphere. The corona is expected to be hotter at low altitudes than the outer layers. Therefore, the coronal structure suggests that the high temperature blackbody component be associated with a smaller volume than the cold component. \citealt{vanderhorst2012} report a strong negative correlation between emission area and blackbody temperature indicating that if the underlying emission mechanism is quasi-thermal with gradually changing temperatures and if the dual blackbody is a well approximation of the continuum gradient as expected, the relationship between temperature and coronal structure is in fact apparent in the spectrum. Modeling the corona with such temperature gradient is a hard task especially in the hotter zone due to its anisotropic structure, the intense magnetic field, the twisted magnetosphere geometry and polarization-dependence of the scattering process. Temporal and spectral studies on broad energy ranges as a result help get a better view of the underlying structure as well as to distinguish between non-thermal and thermal models as in our investigations which suggest a non-thermal model describes spectra best for majority of bursts. 

Although the sum of two blackbody functions is commonly used to describe magnetar burst spectra, it was shown that the spectrum of photon flux per unit energy band may be flat at energies lower than the temperature when the magnetic field is not too high (B $<$ \(10^{15}\) G). In this model, the spectrum of photons escaping the bubble formed during the burst are considered. The emergent spectrum was shown by \citealt{lyub2002} to be close to the blackbody spectrum although the observed radiation within the bubble comes from photons with different temperatures throughout the bubble. In a hot, optically thick bubble in a strong magnetic field, the photon energy is well below the excitation energy of the first Landau level and Compton scattering dominates. Considering SGR burst bubbles in such a scenario, photons with ordinary orthogonal (O-mode) polarization will go under much more scatterings than the extraordinary linearly polarized (E-mode) photons whose cross-sections are strongly hindered (\citealt{lyub2002}). Because of this dependence of scatterings on radiation cross-section which in turn depend on frequency in a strong magnetic field, at low energies the burst spectrum alters from the blackbody spectrum and may be observed as flat.

In our investigations, the LB+LB model emerged as the most frequently chosen test model. That is, compared to the other two models with the same degrees of freedom (namely BB+PO and BB+BB), the LB+LB model shows the best test statistics for 48 out of 77 bursts for SGR 1806$-$20, 16 out of 21 for SGR 1900+14 and 2 out of 4 for SGR J1550$-$5418 when $<$ 50\% parameter error constraints are enforced. This is suggestive that the modified blackbody scheme where a frequency dependent scattering cross-section distorts the blackbody emission at low energies explains burst emission mechanisms better than a thermal emission resulting from short lasting electron-positron pairs through a gradient of temperature at least on time-averaged burst spectra. 

\subsection{Spectral Correlations}
We have also explored whether there exists any correlations between magnetar burst spectral parameters. We investigated spectral correlations of COMPT model because our simulation results show that COMPT model describes majority of well-restricted bursts better described by COMPT model and BB+BB model parameters for comparison purposes.  In the correlation analyses, we chose unsaturated bursts that have well-restricted ($<$ 50 $\%$ of the parameter) parameter and flux errors for the given model fit (COMPT model for Section 6.1.1 and BB+BB model for Section 6.1.2) out of all bursts included in spectral analysis. For the COMPT model, our correlation analysis consists almost entirely of SGR 1806$-$20 and SGR 1900+14 bursts (total of 63) except for one SGR J1550$-$5418 bursts due to error restrictions. 

\subsubsection{COMPT Model}
For the COMPT model, we checked for correlations including only unsaturated bursts that yield well-constrained  ($<$ 50 $\%$ of the parameter) errors for all parameters and flux. For these bursts, we find a weak positive correlation (Spearman's rank correlation coefficient, $\rho$ = 0.54, chance probability = $ 4.05 \times 10^{-6}$) between photon index (defined as $\alpha$ in Equation 1) and total fluence (see Figure 8, right panel). Note that we do not find any significant correlation between $E_{peak}$ and total flux ($\rho$ = 0.22, chance probability = $0.075$).  We find that although the correlation between photon index and total flux is not significant ($\rho$= 0.43, chance probability = $4.06 \times 10^{-4}$), there is a positive lower bound trend on photon index with respect to total flux (see Figure 8, left panel). That is, for the lower-flux bursts, photon index spans a wider range ($-1.1$ to 1.3) while the photon index range is higher for bursts with higher total flux. We take that the negative photon index ($-\alpha$) is a good indicator of the hardness of burst spectra since $E_{peak}$ values of these bursts are narrowly distributed (a gaussian fit on $E_{peak}$ distribution yields $\sigma = 10.6 \pm 0.7$ for SGR 1806$-$20 and  $\sigma = 8.8 \pm 1.1$ for SGR 1900+14, see also Figures 4 and 6) and since $E_{peak}$ does not significantly depend on burst flux. Thus, in low flux ranges, we find that the bursts can be spectrally much harder, while with increasing flux, bursts are confined to softer spectra with a clear lower bound (indicated with the dashed line in Figure 8, left panel) on photon index. Our results spectrally confirm the results of \citealt{gogus2001}, in which they found that high fluence SGR 1806$-$20 and SGR 1900+14 bursts tend to be softer (i.e., anti-correlation between hardness and energy fluence) where hardness ratio is defined as the ratio of photon counts in 10$-$60 keV to 2$-$10 keV bands. Additionally, \citealt{vanderhorst2012} also found an anti-correlation between hardness and fluence for SGR J1550$-$5418 bursts inferred from the anti-correlation they find between $E_{peak}$ and burst fluence from broadband (8$-$200 keV) spectral analysis results. On the other hand, \citealt{gavriil2004} noted a positive correlation between hardness and fluence for the bursts from another magnetar candidate, AXP 1E 2259+586. Overall, we confirm the findings of \citealt{gogus2011} and see that SGR 1806$-$20 and SGR 1900+14 bursts have opposite trends between hardness and fluence in similar fluence ranges ($2.4 \times 10^{-9} - 6.3 \times 10^{-8} erg/cm^2$) to those of AXP 1E 2259+586 (\citealt{gavriil2004}) and similar trends to those of SGR J1550$-$5418 (\citealt{vanderhorst2012}).  We suggest this relationship may be a distinctive characteristic between AXP and SGR bursts.

\begin{figure}[!h] \label{fig:phoindcorrel}
    \plottwo{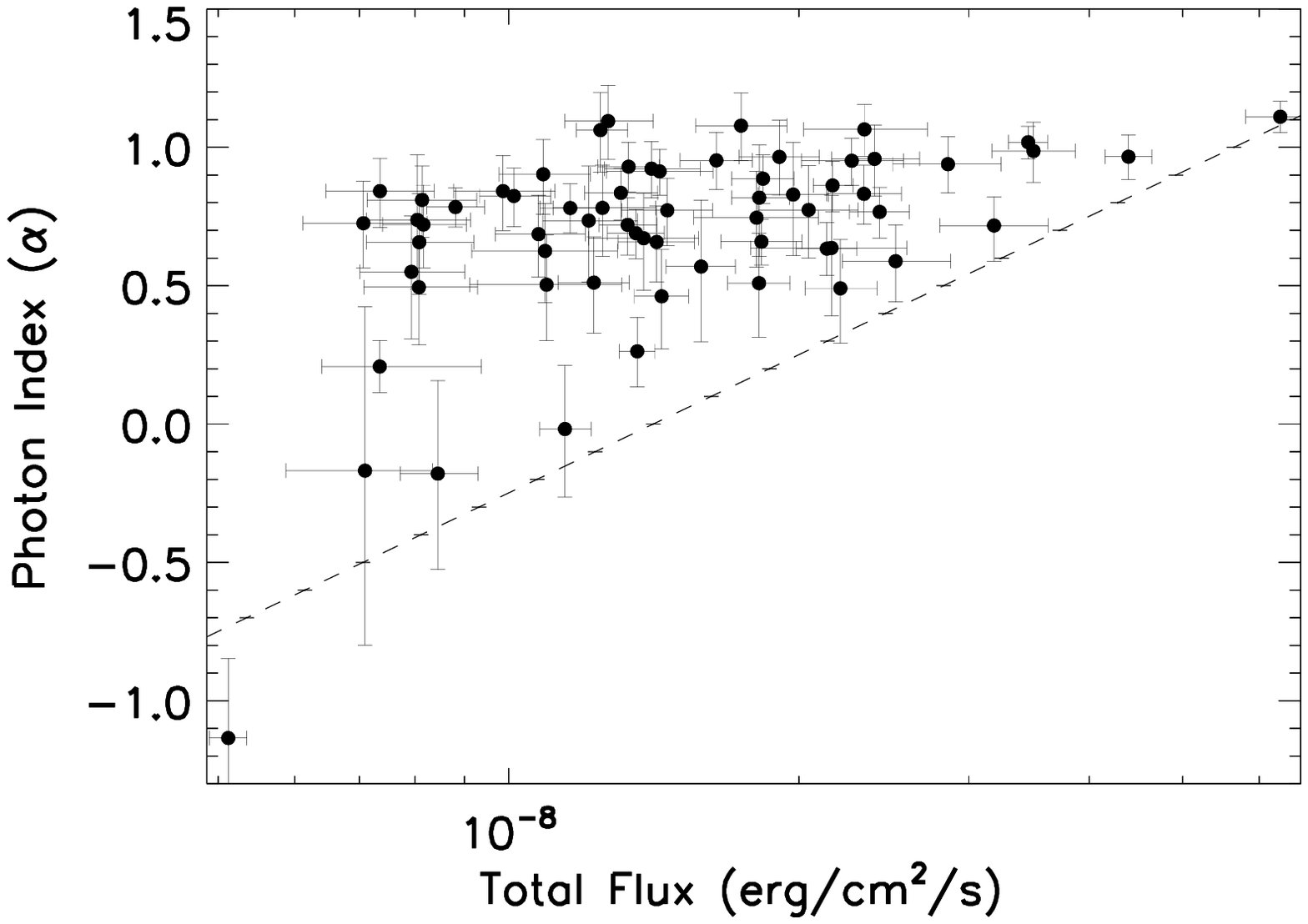}{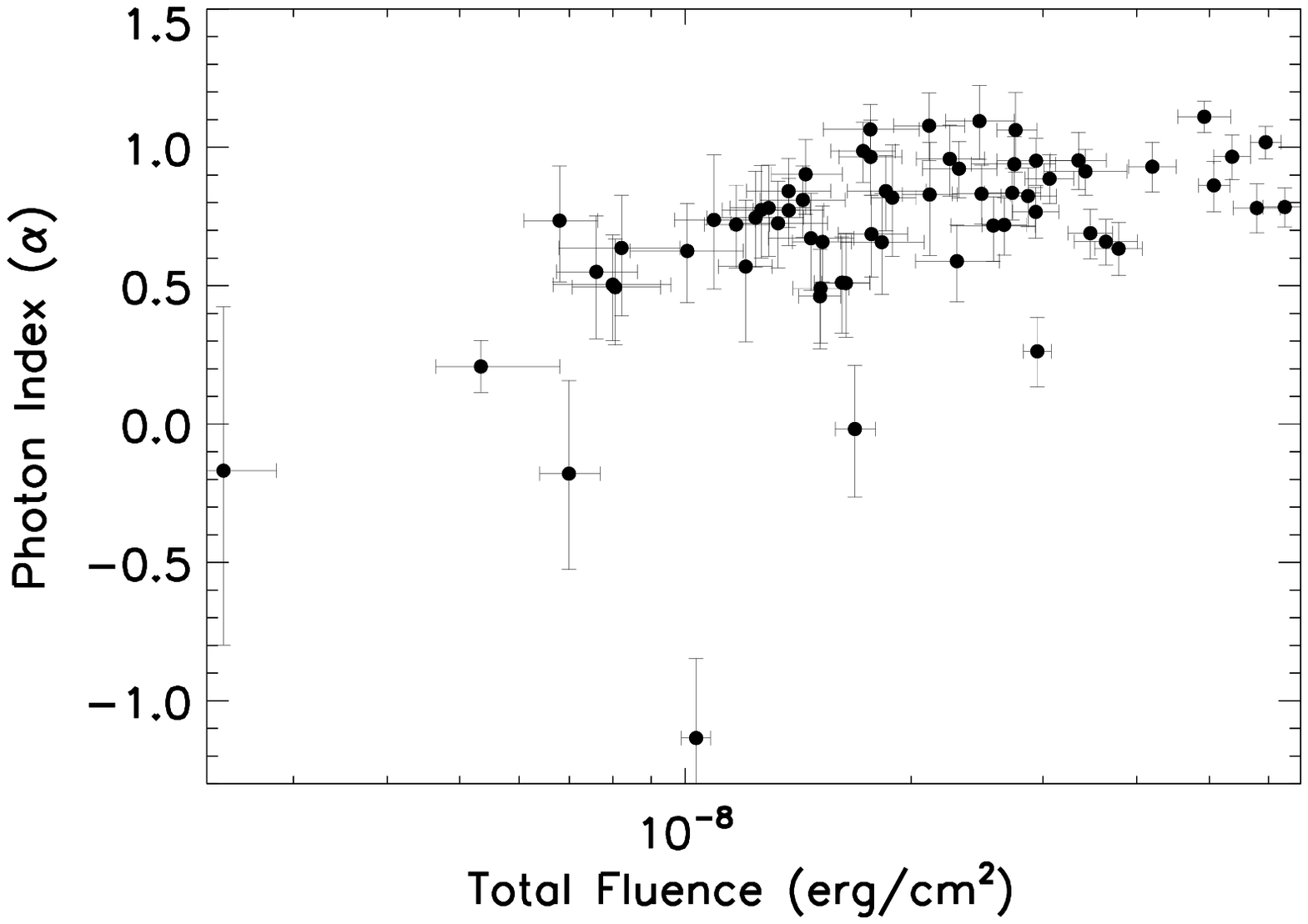}
     \caption{\textit{Left panel} COMPT model photon index vs. total burst flux for unsaturated bursts with well-restricted parameters. Dashed line is drawn to visualize the lower bound trend on photon index. \textit{Right panel}  COMPT model photon index vs. total burst fluence}
\end{figure}

\subsubsection{BB+BB Model}
In the thermal emission viewpoint, a temperature gradient throughout the emitting surface is assumed. This gradient from the burst ignition point has been represented in previous studies by two (hot and cooler) blackbody components (see e.g., \citealt{lin2012}). We also checked the relationship between thermal emission areas and temperatures between of the hot and cold blackbody components, expressed as 
\begin{equation}
 R^2 = FD^2/\sigma T^4
\end{equation}
where \textit{R} is the radius of the thermal emitting region, \textit{D} is the distance to source and \textit{F} is the average total flux per event, and \textit{T} is the temperature. Here, we used the distances of 5 kpc for SGR J1550-5418 (\citealt{tiengo2010}), 12.5 kpc for SGR 1900+14 (\citealt{davies2009}) and 8.7 kpc for SGR 1806-20 (\citealt{bibby2008}). We have found a significant anti-correlation between thermal emission area and temperature for both blackbody components.
 
We present in the upper left panel of Figure 9, the R$^2$ (km$^2$) vs. kT (keV) trend for all unsaturated bursts with well-constrained parameters from all three sources. The dashed lines represent the \(R^2 \propto T^{-3}\) trend proposed in previous studies (e.g. \citealt{israel2008}; \citealt{younes2014}) and the solid line represents the \(R^2 \propto T^{-4}\) trend that is expected from blackbody emission. These trends are drawn for comparison only. The cooler blackbody components ($kT_1$, shown in black) are well separated from the hot blackbody ($kT_2$, shown in red) components for all sources, similar to the results by \citealt{vanderhorst2012} and \citealt{lin2012} for SGR J1550$-$5418 and \citealt{israel2008} for SGR 1900+14. This verifies the thermal emission model that starts from the hot and narrower ignition point and extends to a wider area where it is cooled down (i.e. hotter corona at lower altitudes than the outer layers). To check the extent of this verification, we employed a Spearman rank order correlation test on the emission area vs. temperature trend, and found a correlation coefficient of $-$0.933 with a chance probability close to nil using all bursts. The correlation coefficients are $-$0.953, $-$0.962, $-$0.942 when the Spearman Correlation test is employed individually for SGR J1550$-$5418, SGR 1900+14 and SGR 1806$-$20, respectively (with chance probabilities $\sim$ 0).  We then employed single power law and broken power law fits on emission area vs. blackbody temperature data. For SGR 1900+14 and SGR 1806$-$20 bursts, the reduced $\chi^2$ values for the broken power law (Table 9, column 6) fits are less than single power law reduced $\chi^2$ values (Table 9, column 8). This suggest the emission area vs. temperature behavior significantly differs between the cool and hot blackbody components. Moreover, the hot blackbody component is associated with a narrower emission area than the cool component. Thus, we re-confirm that our findings on this strong anti-correlation between temperature and emission area verify the thermal emission model approximated by a sum of two blackbodies where emission starts from a hot and narrower ignition point within the bubble and extends to a cooler outer layer associated with higher area. 

To check whether similar relationships hold for different burst intensity levels, we grouped the bursts into three intervals based on their total flux and employed a single and broken power law fit for each group in the same flux interval individually. We show a color-coded scatterplot of different flux values for SGR 1806$-$20 in the upper right panel of Figure 9. The dashed lines represent the best fit broken power law trends. The color coded arrows mark the power law break energy for each intensity groups. We present the complete fit results in Table 9. Note that kT uncertainties are  presented in Table 9 and Figures 9b,c and d. R$^2$ uncertainties are propagated ignoring covariance terms and therefore are overestimated.

We find that as burst intensity increases, power law index and power law break (in kT) tend to increase for all three groups. We note that the power law trends between the temperature and emitting area are similar for cool and hot blackbody components at flux values below 10$^{-7.9}$ $erg/cm^2/s$; namely, a single power law represents the trends for both cooler and hot components. For the highest flux group, the power law trends for the cool and hot components differ significantly. In line with this result, the difference between broken power law indexes as well as reduced $\chi^2$ values of single and broken power law fits increase with increasing burst intensity, indicating that the temperature and emission are relation between the two blackbody components differ more with increasing burst intensity.

\begin{figure}[!h] \label{fig:bbbbcorrel}
    \plottwo{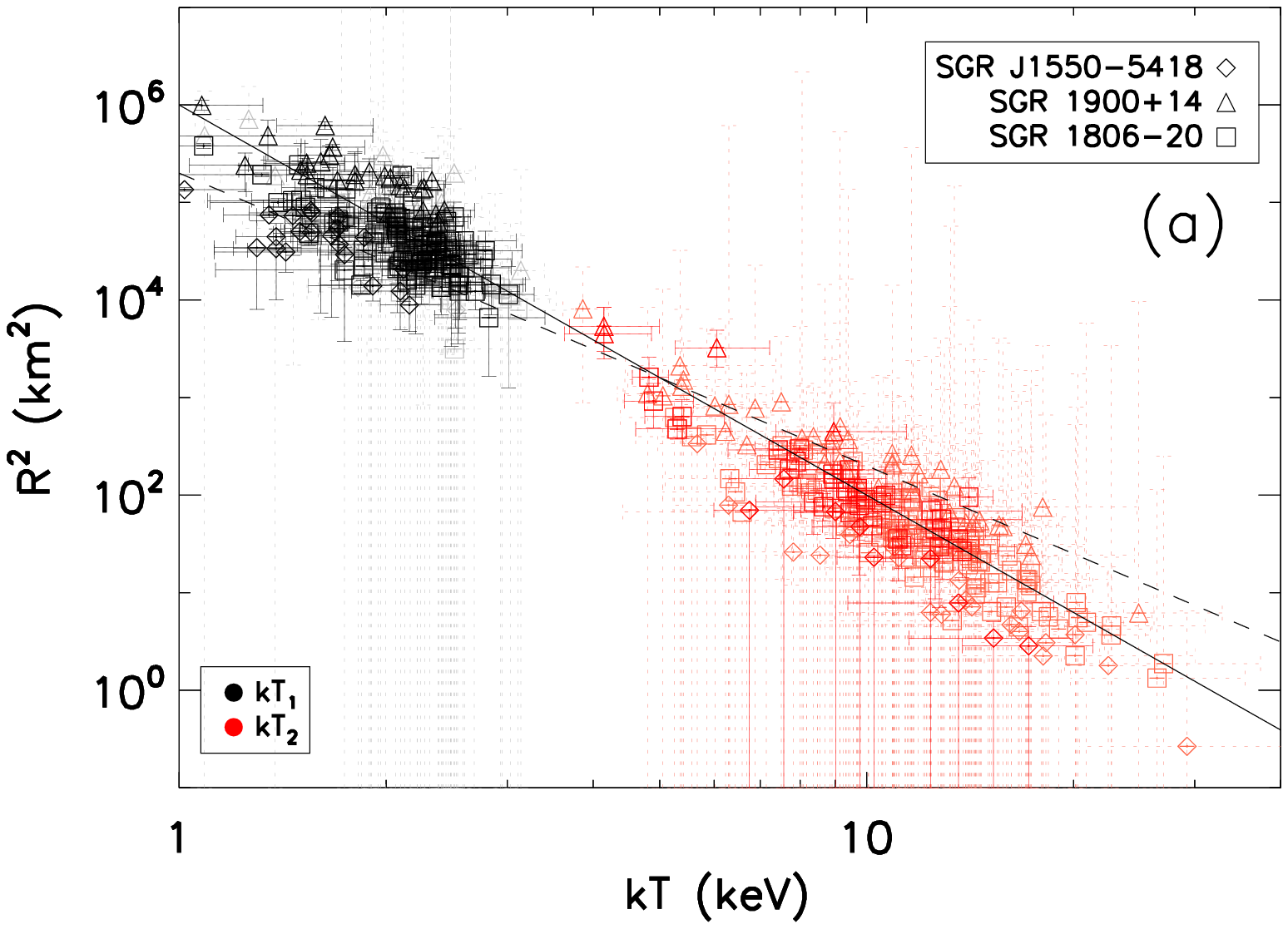}{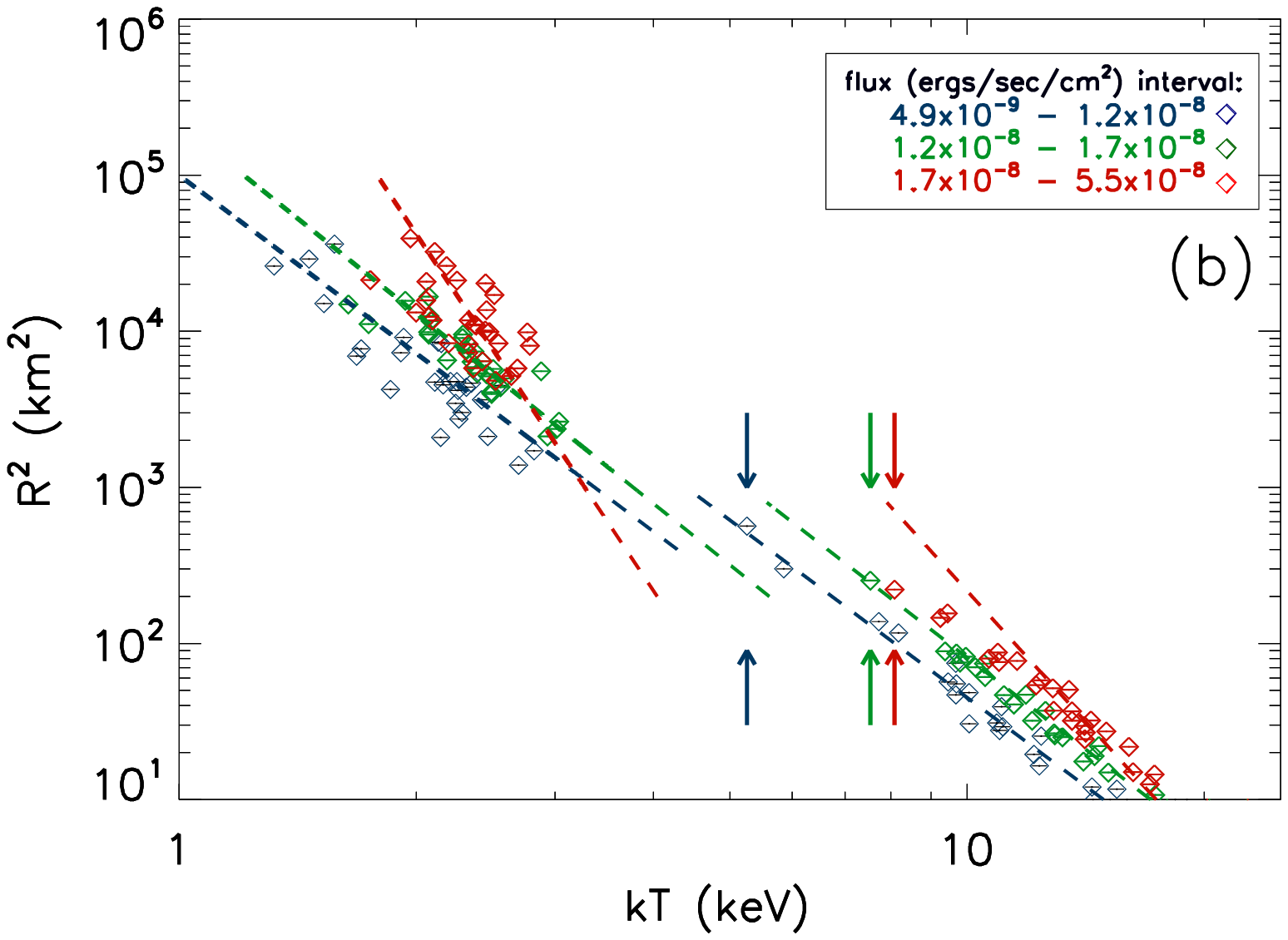}
    \plottwo{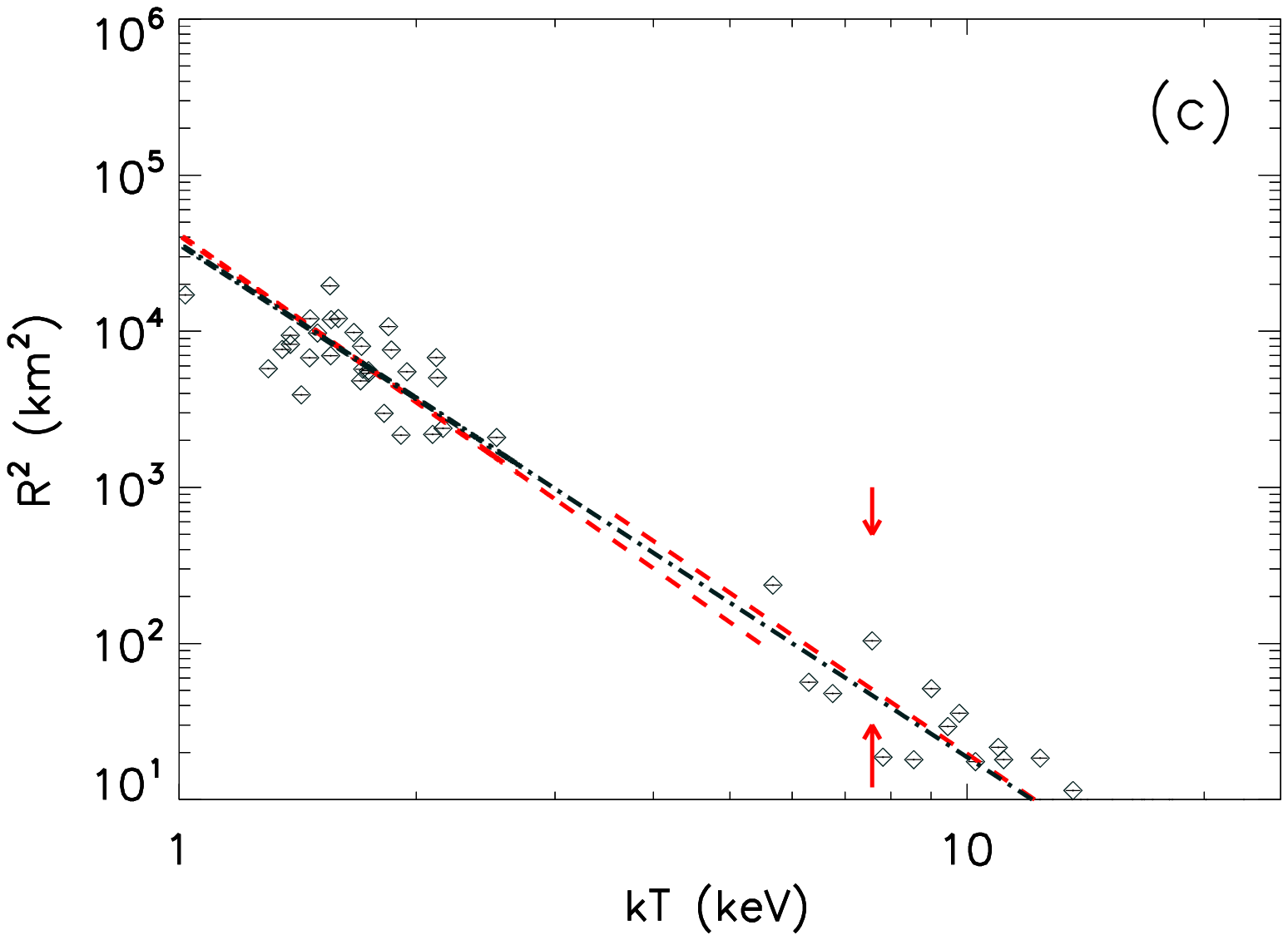}{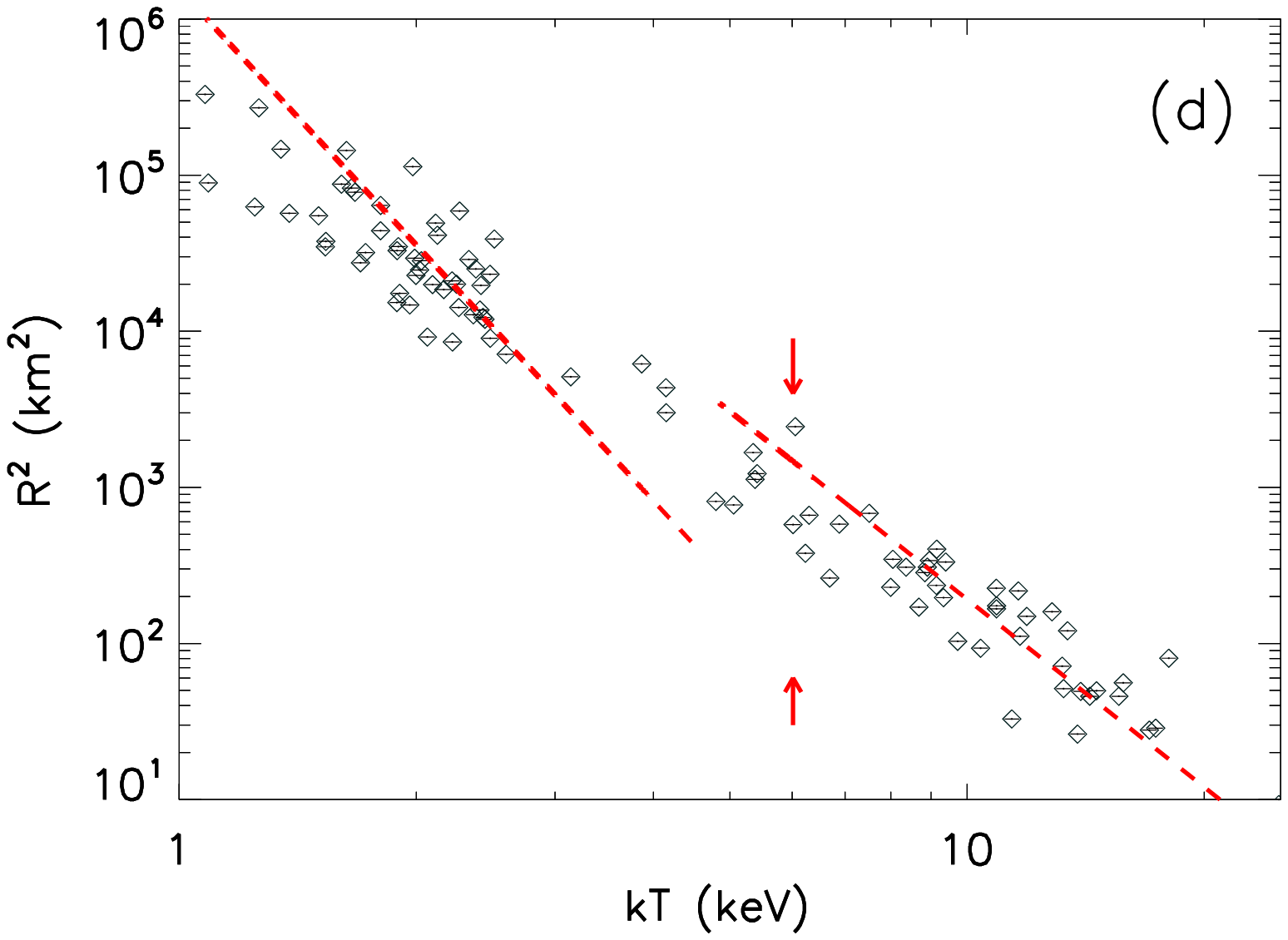}  
 \caption{\textit{(a)} SGR J1550$-$5418, SGR 1900+14 and SGR 1806$-$20 emission area vs. hot and cold blackbody temperatures. \(R^2 \propto T^{-3}\) and \(R^2 \propto T^{-4}\) are drawn with dashed and solid lines respectively for comparison only. Data points with lighter colors are associated with large propagated uncertainty in $R^2$ (i.e. $\delta R^2/R^2 \geq 1$). Note that $R^2$ errors are propagated neglecting the covariance terms and are thus overestimated. \textit{(b)} SGR 1806$-$20 emission area vs. hot and cold blackbody temperatures grouped by total flux values with corresponding broken power law fits. Break index in kT space (keV) are shown with color-coded arrows. \textit{(c)} SGR J1550$-$5418 emission area vs. hot and cold blackbody temperatures with broken power law (red dashed line) and linear model (black dashed line) fits. Arrows represent break energy in kT space (keV). \textit{(d)} SGR 1900+14 emission area vs. hot and cold blackbody temperatures with broken power law fit shown in dashed line. Break index in kT space (keV) is shown with the arrows. Note that only non-saturated bursts are included in all correlation analyses and kT uncertainties are included in the fits}.
\end{figure}

\begin{deluxetable}{cccccccc}\label{tab:brokenpl}
\tablecaption{Single and Broken Power Law Fit Results with Corresponding Flux Intervals \label{tab:br_pl_1806}}
\tablewidth{0pt}
\tablehead{
\multicolumn{2}{c|}{}  & \multicolumn{4}{c|}{Broken Power law} & \multicolumn{2}{c|}{Single Power law} \\
\colhead{Source} & \colhead{ \(log_{10}\) Flux} & \colhead{Low-Energy} & \colhead{High Energy} &
\colhead{\(kT_{break}\)} & \colhead{\( \chi^2 /DOF \)} & {Index} &   \colhead{\( \chi^2 /DOF \)} \\
  & \colhead{Interval (\(erg/cm^2/s\))} & \colhead{Index} & \colhead{Index} & &\colhead{ }   & & \colhead{ }  \\
}
\startdata
SGR 1806$-$20 & $-$8.31, $-$7.91 & $-$3.79 $\pm$ 0.03 & $-$3.78 $\pm$ 0.03 & 5.26 & 0.93 & -3.2 $\pm$ 0.08 & 0.98 \\ 
SGR 1806$-$20 & $-$7.91, $-$7.77 & $-$4.04 $\pm$ 0.03 & $-$3.91 $\pm$ 0.04 & 7.54 & 0.4 & -3.23 $\pm$ 0.07 & 0.53 \\ 
SGR 1806$-$20 & $-$7.77, $-$7.26 & $-$7.59 $\pm$ 0.02 & $-$5.57 $\pm$ 0.04 & 8.08 & 0.87 & -3.42 $\pm$ 0.06 & 2.15\\ 
SGR 1900+14 & $-$8.4, $-$7.16 & $-$5.44 $\pm$ 0.01 & $-$4.00 $\pm$ 0.04 & 6.01 & 2.12 & -3.6 $\pm$ 0.08 & 2.84\\ 
SGR J1550$-$5418 & $-$8.43, $-$7.58 & $-$3.55 $\pm$ 0.03 & $-$3.43 $\pm$ 0.03 & 7.58 & 0.92 & -3.29 $\pm$ 0.07 & 0.88\\ 
\enddata
\end{deluxetable}

We could not split the burst sample of SGR J1550$-$5418 and SGR 1900+14 into intensity groups since these sources have much fewer unsaturated bursts with well-constrained parameters. For SGR J1550$-$5418 bursts (see Figure 9, lower left panel), the power indices of the low and high temperature components are very similar, suggesting a single power law trend could represent the whole sample. A single power law fit on SGR J1550$-$5418 yields $R^2 \sim T^{-3.3 \pm 0.07}$ with a reduced $\chi^2$ of 0.88. Note that bursts from this magnetar have flux values less than 10$^{-7.58}$ $erg/cm^2/s$. In the case of SGR 1900+14 bursts (Figure 9, lower right panel), we find a significant change of the power law trends for two blackbody components, in line with the case of higher intensity bursts from SGR 1806$-$20. Note that on average SGR 1900+14 bursts have higher flux values than that of SGR J1550$-$5418.

These relations between emission area and temperature significantly differ from the relations discussed in \citealt{younes2014} where the emitting area decreases more with increasing temperature for the hot blackbody component. However, we note that the flux range analyzed by \citealt{younes2014} for SGR J1550$-$5418 (above $10^{-6.5}$ $erg/cm^2/s$) is much higher from the flux range covered in our investigation ($\sim 10^{-9}-10^{-8}$ $erg/cm^2/s$). It is possible that the area vs. \textit{kT} behavior of the two blackbody components differ more significantly with increasing flux than previously discussed or that the relationship differs between sources or burst episodes. It remains likely that the trend between lower and upper power law indices show opposite behaviors depending on burst intensity. We suggest that the power law trends of the cool and hot components are the same in the flux regime $\sim 10^{-8}$ $erg/cm^2/s$. Above this flux level, the cooler component shows more decrease in emission area with increasing temperature than that of the hot component (as in the high intensity case of SGR 1806$-$20 (Figure 9, upper right panel)). In the much higher flux regime (above $\sim 10^{-7}$ $erg/cm^2/s$), an opposite behavior takes place, as presented by \citealt{younes2014}; namely, the emission area of the hot component drops more than that of the cool component with temperature.

\section{Concluding Remarks}

We presented our time-averaged spectral analysis results of a total of 388 bursts from SGR J1550$-$5418, SGR 1900+14 and SGR 1806$-$20 as machine readable tables in this paper and as a database in a companion web-catalog at \url{http://magnetars.sabanciuniv.edu}. Our spectral analysis show BB+BB, BB+PO, LB+LB and COMPT models all provide acceptable fits at similar levels. We further conducted numerical simulations to contrain the best-fitting model. Based on the simulation results, we suggest COMPT provides significantly better fits with sufficiently constrained parameters. We suggest the inherent emission mechanism is likely non-thermal or at least not purely thermal for majority of bursts included in our study within 2-250 keV range. It is important to note that since our analysis covers time-averaged spectra only, these results may not fully represent instantaneous burst properties. Excluding COMPT fit results, we find that LB+LB model, which is employed in SGR spectral analysis for the first time here, describes majority of bursts with well-restricted parameters better than BB+BB and BB+PO models.

We find that the photon index is positively correlated with fluence and has an increasing lower bound trend with respected to burst flux, suggesting that bursts have a decreasing upper bound limit on hardness  with respect to increasing flux. This behavior is similar to the previously reported anti-correlation between hardness and fluence for SGR J1550$-$5418 bursts, and confirms the anti-correlation between hardness and fluence for SGR 1806$-$20 and SGR 1900+14 bursts. Since it was shown that AXP 1E 2259+586 bursts show an opposite trend (positive correlation between burst hardness and fluence), we suggest that the relationship between hardness and burst fluence is a distinctive behavior between AXP and SGR bursts.

We confirm a significant anti-correlation between blackbody temperatures (hot and cold) and burst emission areas. Overall, emission area decreases with increasing blackbody temperature in all cases, verifying the thermal emission model from a hot bubble where emission radiates from a hot ignition point to a colder and wider emission area. Examining the same relation in different flux intervals of SGR 1806$-$20 bursts, we found that above the flux regime of  \(\sim 10^{-8}\) erg cm$^{-2}$ s$^{-1}$, the emission area decreases more rapidly with respect to the temperature for the cooler blackbody.  This is in contrast to what have been reported previously (e.g. \citealt{younes2014}). We suggest area vs. kT behaviors of the two blackbody components may differ more significantly with increasing flux than previously discussed. It is also possible that the trend between lower and upper power law indices for the cool and hot blackbody components show opposite behaviors at different burst intensities.

\acknowledgments

D.K., Y.K. and S.S.M acknowledge support from the Scientific and Technological Research Council of Turkey (T\"UB\.ITAK, grant no: 113R031)


\begin{thebibliography}{}
\bibitem[Beloborodov \& Thompson(2007a)]{beloborodovthompson2007a} Beloborodov, A.~M., Thompson, C.\ 2007a, \apj, 657, 967
\bibitem[Beloborodov \& Thompson(2007b)]{beloborodovthompson2007b} Beloborodov, A.~M., Thompson, C.\ 2007b, \apss, 308, 631
\bibitem[Bibby et al.(2008)]{bibby2008}  Bibby, J.~L.,  Crowther, L.~A., Furness, J.~P., et al.\ 2008, \mnras, 386, L23
\bibitem[Davies et al.(2009)]{davies2009}  Davies, B.,  Figer, D.~F., Kudritzki, R.~P., et al.\ 2009, \apj, 707, 844
\bibitem[Feroci et al.(2004)]{feroci2004} Feroci, M., Caliandro, G.~A., Massaro, E., et al.\ 2004, \apj, 612, 408
\bibitem[Gavriil et al.(2004)]{gavriil2004} Gavriil, F.~P., Kaspi, V.~M., \& Woods, P.~M.\ 2004, \apj, 607, 959 
\bibitem[Gogus et al.(2007)]{gogus2007} Gogus, E., Woods, P.~M., Kouveliotou, C., et al.\ 2007, Progress of Theoretical Physics Supplement, 169, 12
\bibitem[Gogus et al.(2011)]{gogus2011} Gogus, E., Guver, T., Ozel, F., et al.\ 2007, \apj, 728, 160 
\bibitem[Gogus et al.(2001)]{gogus2001} Gogus, E., Kouveliotou, C., Woods, P.~M., et al.\ 2001, \apj, 558, 228 
\bibitem[Halpern et al.(2008)]{halpern2008} Halpern, J.~P., Gotthelf, E.~V., Reynolds, J., et al.\ 2008, \apj, 676, 1178
\bibitem[Israel et al.(2008)]{israel2008} Israel, G.~L., Romano, P., Mangano, V., et al.\ 2008, \apj, 685, 1114
\bibitem[Israel et al.(2010)]{israel2010} Israel, G.~L., Esposito, P., Rea, N., et al.\ 2010, \mnras, 408, 1387
\bibitem[Jahoda et al.(2006)]{jahoda2006} Jahoda, K., Markwardt, C.~B., Radeva, Y., et al.\ 2006, \apjs, 163, 401
\bibitem[Kaspi \& Beloborodov(2017)]{kaspi2017} Kaspi, V.~M., Beloborodov, A.\ 2017, ArXiv e-prints, arXiv:1703.00068
\bibitem[Lin et al.(2013)]{lin2013} Lin, L., Gogus, E., Kaneko, Y., et al.\ 2013, \apj, 778, 105 
\bibitem[Lin et al.(2011)]{lin2011} Lin, L.,Kouveliotou, C., Baring, M.~G., et al.\ 2011, \apj, 739, 87
\bibitem[Lin et al.(2012)]{lin2012} Lin, L., Gogus, E., Baring, M.~G., et al.\ 2012, \apj, 756, 54
\bibitem[Lyubarsky(2002)]{lyub2002} Lyubarsky, Y.~E. 2002, \mnras, 332, 199
\bibitem[Lyutikov(2003)]{lyutikov2003} Lyutikov, M. 2003, \mnras,  346, 540
\bibitem[Manchester et al.(2005)]{manchestercatalogue} Manchester, R.~N., Hobbs, G.,B., Teoh, A., et al.\ 2005, \aj, 129, 1993
\bibitem[Mereghetti et al.(2015)]{mereghetti2015} Mereghetti, S., Pons, J.~A., \& Melatos, A. \ 2015, \ssr, 191, 315
\bibitem[Nobili et al.(2008)]{nobili2008} Nobili, L., Turolla, R., \& Zane, S. \ 2008, \mnras, 389, 989
\bibitem[Olive et al.(2003)]{olive2003} Olive, J.~F., Hurley, K., Dezalay, J.~P., et al.\ 2003,  American Institute of Physics Conference Series, Vol. 662, Gamma-Ray Burst and Afterglow Astronomy 2001: A
Workshop Celebrating the First Year of the HETE Mission, ed. G.~R. Ricker \& R.~K. Vanderspek, 82?87
\bibitem[Scargle et al.(2003)]{scargle2003} Scargle, J.~D., Norris, J.~P., Jackson, B., et al.\ 2003, \apj,764, 167
\bibitem[Thompson \& Beloborodov(2005)]{thompsonbeloborodov2005} Thompson, C., Beloborodov, A.~M., \ 2005, \apj, 634, 565
\bibitem[Thompson \& Duncan(1995)]{thompsonandduncan95} Thompson, C., Duncan, R.~C., \ 1995, \mnras, 275, 255
\bibitem[Thompson et al.(2002)]{thompsonlyutikovkulkarni2002} Thompson, C., Lyutikov, M. \& Kulkarni, S.~R.\ 2002, \apj, 574, 332
\bibitem[Tiengo et al.(2010)]{tiengo2010} Tiengo, A., Vianello, G., Esposito, P., et al.\ 2010, \apj, 710, 227
\bibitem[Turolla et al.(2015)]{turolla2015} Turolla, R., Zane, S. \& Watts, A.~L.\ 2015, Reports on Progress in Physics, 78, 116901
\bibitem[van der Horst et al.(2012)]{vanderhorst2012} van der Horst, A.~J., Kouveliotou, C., Gorgone, N.~M., et al.\ 2012, \apj, 749, 122
\bibitem[Woods et al.(2007)]{woods2007}  Woods, P.~M., Kouveliotou, C., Finger, M.~H., et al.\ 2007, \apj, 654, 470
\bibitem[Younes et al.(2014)]{younes2014}  Younes, G., Kouveliotou, C., van der Horst, A.~J., et al.\ 2014, \apj, 785, 52
\end{thebibliography}
\end{document}